\documentclass[twocolumn]{aastex62}

\usepackage{multirow}
\usepackage{amsmath}

\graphicspath{{./}{figures/}}

\received{January 1, 2018}
\revised{January 7, 2018}
\accepted{\today}
\submitjournal{ApJ}

\shorttitle{Merger Classifications with Deep Learning}
\shortauthors{Ferreira et al.}
\begin{document}

\title{Galaxy Merger Rates up to $z\sim3$ using a Bayesian Deep Learning Model --- \\ A Major-Merger classifier using IllustrisTNG Simulation data}

\correspondingauthor{Leonardo Ferreira}
\email{leonardo.ferreira@nottingham.ac.uk, conselice@gmail.com}

\author[0000-0002-8919-079X]{Leonardo Ferreira}
\affil{University of Nottingham, School of Physics \& Astronomy, Nottingham NG7 2RD, UK}

\author{Christopher J. Conselice}
\affil{University of Nottingham, School of Physics \& Astronomy, Nottingham NG7 2RD, UK}
\author{Kenneth Duncan}
\affil{Leiden Observatory, Leiden University, PO Box 9513, NL-2300 RA Leiden, the Netherlands}
\affil{SUPA, Institute for Astronomy, Royal Observatory, Blackford Hill, Edinburgh, EH9 3HJ, UK}

\author{Ting-Yun Cheng}
\affil{University of Nottingham, School of Physics \& Astronomy, Nottingham NG7 2RD, UK}
\author{Alex Griffiths}
\affil{University of Nottingham, School of Physics \& Astronomy, Nottingham NG7 2RD, UK}
\author{Amy Whitney}
\affil{University of Nottingham, School of Physics \& Astronomy, Nottingham NG7 2RD, UK}



\begin{abstract}
Merging is potentially the dominate process in galaxy formation, yet there is still debate about its history over cosmic time.  To address this we classify major mergers and measure galaxy merger rates up to $z \sim 3$ in all five CANDELS fields (UDS, EGS, GOODS-S, GOODS-N, COSMOS) using deep learning convolutional neural networks (CNNs) trained with simulated galaxies from the IllustrisTNG cosmological simulation. The deep learning architecture used is objectively selected by a Bayesian Optmization process over the range of possible hyperparameters. We show that our model can achieve 90\% accuracy when classifying mergers from the simulation, and has the additional feature of separating mergers before the infall of stellar masses from post mergers. We compare our machine learning classifications on CANDELS galaxies and compare with visual merger classifications from \cite{Kartaltepe2015}, and show that they are broadly consistent. We finish by demonstrating that our model is capable of measuring galaxy merger rates, $\mathcal{R}$, that are consistent with results found for CANDELS galaxies using close pairs statistics, with $\mathcal{R}(z)= 0.02\pm0.004 \times (1+z)^{2.76\pm0.21}$.  This is the first general agreement between major mergers measured using pairs and structure at $z < 3$. 
\end{abstract}



\section{Introduction}
\label{sec:intro}

Galaxy mergers are an explicit display of the hierarchical assembly of the universe, where galaxies and their dark matter halos merge together to form more massive systems \citep[e.g.][]{2010gfe..book.....M}. Indeed, the rate by which galaxies merge is a consequence of how the universe evolved, and can be used as an observable for the history of mass assembly of galaxies \citep{Conselice2014a}. The understanding of how mass is assembled by galaxies is a very important piece of the galaxy formation and evolution landscape. It is known to happen in two ways: merging \citep{Duncan2019} and through the accretion of gas from the environment, resulting in star formation \citep{Almeida2014}. The contribution of star formation to the mass assembly of galaxies is well measured even to high redshifts, where a peak in star formation rates are observed around $z \sim 2$ \citep{Madau2014}. The contribution from mergers, however, is less straightforward to measure and has some difficulties linked to how we identify merging systems \citep{Conselice2006, Lotz2008, Conselice2014,Man2016}.

Overall, two distinct methods are currently used to find galaxy mergers. One consists of finding close pairs of galaxies that fulfill a maximum separation criteria (both in redshift and angular separation) such that their orbits will dynamically decay with time resulting in a merger event. This is a quite successful approach and enabled merger fractions and rates to be estimated up to $z \sim 6$ \citep[e.g.][]{Mundy2017, Duncan2019}. The second method relies on non-parametric morphological measurements that are robust for finding galaxies with disturbed morphologies, which is a strong suggestion (but not solely) for galaxy merging and interactions. In this case, a suite of measurements, generally the CAS  (Concentration, Asymmetry, Smoothness) and the G-$M_{20}$ systems, are used together to generate a parameter space which serves as a diagnostic tool for galaxy morphological classification \citep{Conselice2003a, Lotz2004}. Some regions of this parameter space are dominated by merging galaxies, which then can be used to  determine if a galaxy is likely a merger or not \citep{Conselice2003a, Lotz2004, Lotz2008}.

Both methods have had success \citep{Conselice2003, Lotz2004, Conselice2009, Mundy2017, Duncan2019}, but they probe galaxy mergers in different ways and rely on different assumptions. For example, in the case of galaxy pairs, merger fractions and rates are measured taking into consideration that the merger event did not happen yet, and may not happen, while the traditional  non-parametric approach is only able to probe around one third of the period of the merger event, when morphologies are disturbed enough to distinguish from normal galaxies \citep[Hubble type galaxies;][]{Conselice2006}. On top of that, it is not only galaxy mergers that populate merger regions of parameter space generated by non-parametric measurements. Other types of galaxies can have signatures that produce similar values, and not all mergers occupy that defined parameter space for the entirety of the merging event. This results in some contamination, generally from star forming galaxies, where star formation regions show themselves as clumpy light in the morphology of the galaxy which can, by eye mimic  the appearance of an ongoing merger. 

Another problem inherent in measuring merger rates is the knowledge of the time-scales involved in the merger event. It is very difficult to infer time-scales from observations, as we are limited to a single snapshot for each observed galaxy, and the merging timescale depends on several dynamical properties of the system \citep{Lotz2008, Conselice2009}. Fortunately, galaxy simulations can be used to estimate such timescales. Not only that, it is also possible to infer timescales attached to each method, for they probe different stages of the merger event \citep{Lotz2008}. Thus, large scale cosmological simulations can be used to estimate the dependence on redshift of merger timescales and visibilities \citep{Snyder2017}. 

This scenario motivates us to develop new methods of finding mergers, and to improve upon current methods. One potential way to make progress in this direction is by using Deep Learning techniques where groups and layers of functions are laid out in a structure inspired by how the neurons in our brain works. In fact, some of these techniques, such as Convolutional Neural Networks, are dedicated to solve computer vision problems \citep[CNNs;][]{Goodfellow-et-al-2016}. For instance, CNNs are widely used in astronomy to tackle several problems, like galaxy morphological classification, segmentation and deblending \citep[e.g.][]{Huertas-Company2018, Reiman2019, Huertas-Company2019, Cheng2018, Martin2019}. 

One of the attempts to detect galaxy mergers with CNNs was done by \cite{Ackermann2018}, where their network was trained with SDSS data labeled with classifications from \cite{Darg2010}. They were able to detect new mergers in the SDSS data that were not originally found by \cite{Darg2010}. This shows that indeed, CNNs are able to learn imaging aspects of merging galaxies. However, any bias in the classifications from \cite{Darg2010} are also incorporated in the model, since galaxies used for training were classified by eye. 

Another experiment was conducted by \cite{Pearson2019}, where galaxy mergers from the EAGLE cosmological simulation \citep{Schaye2015} were used to train a CNN. In cosmological simulations such as this the merger history of all simulation galaxies is available through merger trees generated by Friend-of-Friends methods. This is a potential solution for labelling training data since this represents a ground truth relative to when two galaxies (or more) are merging, in contrast to eyeball classifications that can be uncertain. These authors also conduct cross training experiments, where simulated galaxies are classified with models trained with real galaxies, and the other way around. However, the results from the application of this trained model fails to classify galaxy mergers, even within the simulation. They attribute the performance of the network to the difference between EAGLE galaxies and real galaxies. Their conclusions is that mergers in the simulation have different morphologies from real galaxy mergers. This can be a result of low resolution or low training sample size, since they only use a few thousand galaxies for training.

A different approach was recently employed by \cite{Snyder2019}, where the authors used a combination of non-parametric morphological parameters, random forests, and ensemble learning to create a model which is capable of classifying galaxy mergers using the Illustris simulation \citep{Vogelsberger2014} galaxies as the training sample. This approach however does not use the embedded powerful feature extraction layers present in CNNs and resembles more the classic classification methods in combination with some of the aspects of basic machine learning.

With this background in mind, we further explore how deep learning methods can help us extract more information regarding mergers from imaging data. We do this by training a model with only simulated data labeled with information available from merger trees in cosmological simulations. {\color{black}{This has the potential to avoid biases that emerge from visual classifications, and by leveraging all the potential information deep learning methods provides, we can construct a full probabilistic approach to conduct predictions in real galaxies}}. 

To do this, we construct a sample of galaxies from the IllustrisTNG suite of cosmological simulations \citep{Nelson2019} with their complete merger histories available as a training sample, and then train a CNN to distinguish major mergers from non-merging galaxies with the goal of applying this to The Cosmic Assembly Near-infrared Deep Extragalactic Legacy Survey (CANDELS)  fields \citep{Grogin2011,Koekemoer2011}. We check if our results are consistent with visual classifications from \cite{Kartaltepe2015} and galaxy merger rates from \cite{Duncan2019}.

This paper is organized as follows: in \S \ref{sec:data} we describe how the data from IllustrisTNG was prepared while we elaborate our Deep Learning architecture in \S \ref{sec:methods}. We dedicate \S  \ref{sec:results} to discuss our results both with the simulation data and real data and we summarize the paper in \S \ref{sec:summary}. All transformations and measurements here assume the same cosmological model used by IllustrisTNG, which are consistent with \cite{PlanckCollaboration2018} results that show $\Omega_{\Lambda,0}=0.6911$, $\Omega_{m,0}=0.3089$ and $h=0.6774$. Magnitudes are quoted in the AB system \citep{Oke1983} unless otherwise specified.

\section{Data} \label{sec:data}

Our goal is to develop a major-merger classifier model trained with galaxies from cosmological simulations and explore whether it is capable of carrying out predictions on real galaxies. In these simulations, a galaxy's complete merger history is generally available through merger trees \citep{Rodriguez-Gomez2015}. This approach enables us to use a completely objective way of labelling our training data, bypassing any visual bias that might affect visual classifications, especially in this merger/non-merger classification task that deals with morphological features that can be the result of several processes, not only merging. However, this comes with drawbacks. The resolution of the simulation must be good enough to generate similar morphologies to the ones present in real galaxies. Not only that, but post-processing steps are necessary to mimic the same observational effects and characteristic noise of the data where predictions will be conducted. Thus, it is of utmost importance that the simulation is able to provide enough galaxy numbers for the classification task (i.e tens of thousands), {\color{black} as we expect it to be able to generalize to a different dataset}. We also want to probe galaxies to moderate redshifts ($0 < z \le 3)$ so we can estimate galaxy merger rates using our predictions. 

\subsection{IllustrisTNG}
\label{sec:illustris_data}

All these requirements lead us to the IllustrisTNG project \citep{Nelson2019}, a suite of cosmological, gravo-magnetohydrodynamical simulation runs, ranging within a diverse set of particle resolutions for three comoving simulation boxes of length size, $50, 100, 300 \rm \ Mpc \ h^{-1}$, named TNG50, TNG100 and TNG300, respectively. Each of these simulations probe a different resolution regime, in a trade-off between galaxy numbers and simulation resolution. As we are interested in building a large training sample, we recur to the largest simulation available, TNG300. Within each simulation box there are also different setups, with variations in the number of gas and dark matter particles. We limit ourselves to the highest resolution available in the largest simulation box, namely TNG300-1\footnote{As a comparison, the TNG100-1 simulation has approximately $4.3$ million subfind groups at $z=0$ while TNG300-1 has $14.4$ million. These groups are sets of simulation particles that are bound together by the Sublink algorithm, which in a general sense can represent galaxies.}. 

It is important to note, however, that the physical resolution of TNG300-1 does not perfectly match the CANDELS resolution, especially at higher redshifts. TNG100-1 and TNG50 would provide better resolution matched candidates if the dominant concern was physical resolution. Instead, our choice here was driven by the simulation volume, and the need to have the largest number of galaxies available to train our machine learning. As a way to mitigate potential issues that could come with this resolution mismatch we only use in our analysis massive galaxies with $M_* > 10^{10} M_\odot$ and major mergers in the case of mergers. 

From TNG300-1 we draw two samples: a major-mergers (hereafter \textbf{MM}) only sample and a sample of non-interacting galaxies (hereafter \textbf{NM}). Details on how both samples are selected are described in \S \ref{subsec:major_mergers} and \S \ref{subsec:non_interacting}, respectively. After selecting and creating a sample of clean galaxy images from IllustrisTNG, we need to apply effects to the imaging data to generate realistic galaxy mocks,  this process is described in \S \ref{sec:imaging_data}. For our sample of real galaxies, we choose to use galaxies in all of the CANDELS fields (COSMOS, UDS, GOODS-S, GOODS-N and EGS). How we select galaxies from CANDELS is described in \S \ref{sec:CANDELSFIELDS}.

\subsubsection{Major-Merger (\textbf{MM}) Sample}\label{subsec:major_mergers}

All our samples are selected through available merger trees. First, we limit our exploration to $z \le 3$ (snapshots 99 to 25). {\color{black} As we will later use near-infrared imaging, this redshift limit is applied to ensure that we are not probing rest-frame UV observations}. We limit this work to the near-infrared to mitigate the  effects of dust attenuation, as the IllustrisTNG imaging data used here is not produced by a proper radiative transfer process. As such, it is essential to avoid probing the rest-frame UV of the simulated galaxies where the effects of dust would be extreme. Thus, within our redshift range we expect the impact of dust to increase as our rest-frame wavelength is closer to the UV rest-frame. A full radiative transfer treatment of the images would be necessary to completely avoid this problem. An alternative would be to use longer wavelengths, which will be possible with JWST imaging in the future. However, both solutions are beyond the scope of this paper.

Then, for each galaxy at $z = 0$ (snapshot 99), we climb the merger tree by checking for cases where there is more than one progenitor in a previous snapshot that fulfill the major-merger mass ratio, $\mu$, criteria,

\begin{equation}
    \mu \ge \frac{1}{4},
\end{equation}{}

\noindent and at least one of the progenitors has $M_* \ge 10^{10} \ M_\odot$. If that is the case, we select the snapshot where these criteria are met as the central snapshot of the merger event. This means that this is the snapshot where the sublink algorithm decided that particles from its progenitors became one descendant. However, it is still possible that in the central snapshot such galaxies are still separated by some distance in the sky, but will appear as only one galaxy in snapshots moving forward. With the central snapshot defined, we select all progenitors and descendants within $\pm \  0.3 \ \rm Gyr$ of the central snapshot as mergers as well. By doing so, we are selecting galaxy mergers in different stages of the merger event around a well defined time-scale. Galaxies in this selection window can appear as pairs, disturbed morphologies that indicate recent infall, and also cases where two or more galaxies already merged and little to no disturbance is visible. 

For all selections before the central snapshot, we measure the distance between each progenitor, $D_n$. Here we apply an additional cut by limiting the distance between each pair of galaxies by $D_n < 20 \ \rm kpc \ h^{-1}$. We are only interested in galaxies that are close enough to appear as if they are going to merge in the future. Such distance separation is within the range generally used for close-pair studies \citep[e.g.,][]{Duncan2019}, but we use it in the lower limit so that all pairs of galaxies involved in a merger event can be sampled in the image's field of view used in this work. 

This selection procedure yields $\sim 30,000$ distinct major-merger candidates. The information in each selected object with respect to its central snapshot enables us to also categorize this sample further in different cases of mergers. All selected objects that have redshifts higher or equal to the redshift of the central snapshot are marked as merger candidates before the merger event (hereafter \textbf{BM}) and the cases with redshifts lower than the central snapshot's redshift are considered post-mergers (hereafter \textbf{PM}).

\begin{figure}
    \centering
    \includegraphics[width=0.5\textwidth]{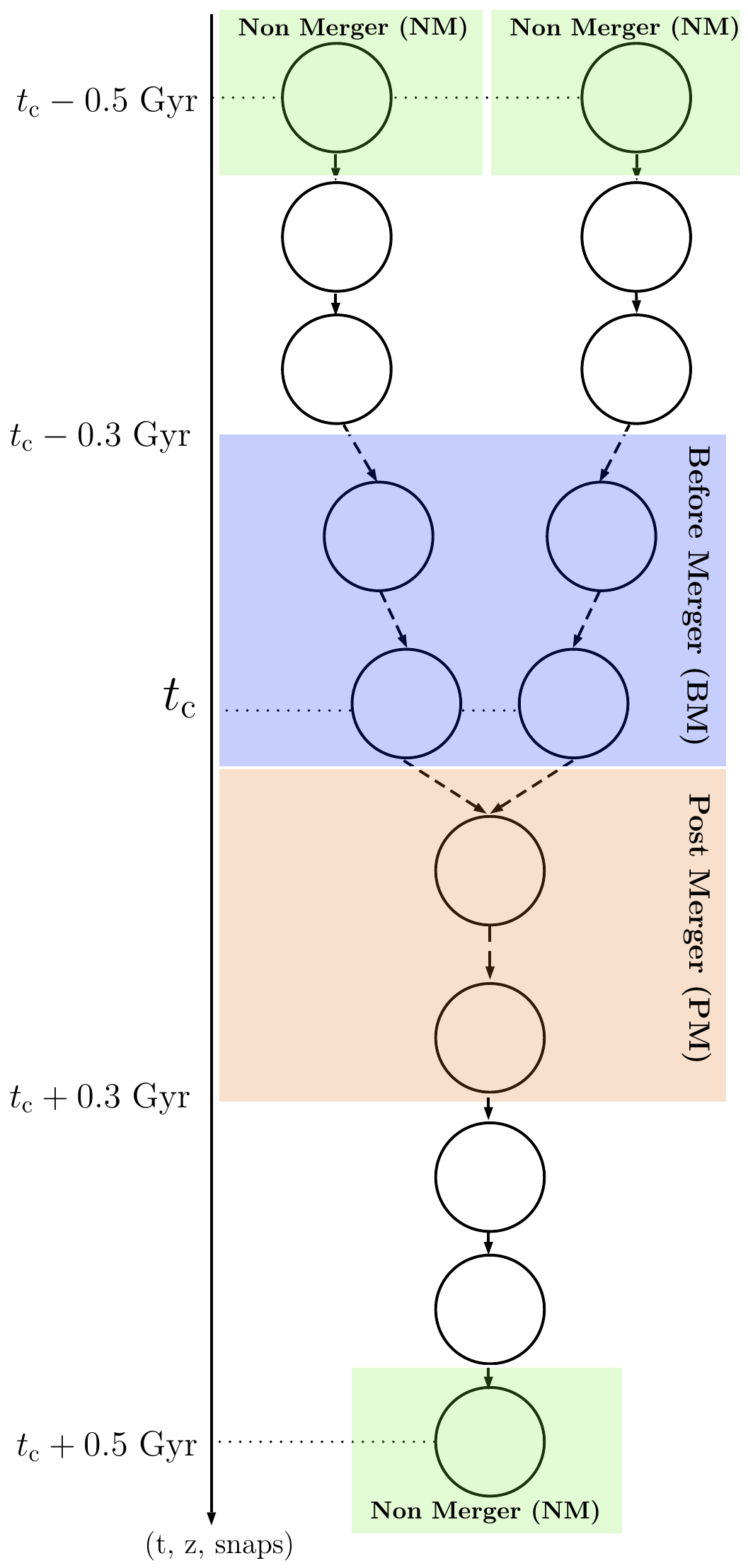}
    \caption{Diagram with a simplified example of two galaxies merging and the resulting label selection for each object and snapshot. Area in blue shows galaxies selected with \textbf{BM} labels, orange represent galaxies with \textbf{PM} labels and in green \textbf{NM}. Both \textbf{BMs} and \textbf{PMs} are selected with our selection timescale, $\tau_{\rm obs} = 0.3 \ \rm Gyr$ , whilst \textbf{NMs} are defined with a longer interval from the central snapshot. Selection windows are drawn based on the central snapshot, $t_c \pm \tau_{\rm obs}$. The \textbf{BM} window include the central snapshot.  }
    \label{fig:mergerdiagram}
\end{figure}

This will not limit our approach towards classifying galaxy mergers only in these two classes, as in \S \ref{subsec:bayesian} we will show that we can still use the prior probability to do a \textbf{MM}/\textbf{NM} classification instead of a \textbf{BM}/\textbf{PM}/\textbf{NM} classification. The only difference when moving from specialized classes to general mergers is using appropriate corresponding observing timescales. It is necessary to use $\tau_{\rm obs} = 0.3 \ \rm Gyr$ when working with \textbf{BM} and \textbf{PM} classes, and $\tau_{\rm obs} = \rm 0.6 \ Gyr$ when working with \textbf{MM} in general, to appropriately reflect our sampling windows. To help with the visualization of our method, we show in Fig. (\ref{fig:mergerdiagram}) a simplified sketch of our selection criteria for two galaxies undergoing a merger.

{}

\subsubsection{Non-Merger (\textbf{NM}) sample}\label{subsec:non_interacting}

A sample of non-mergers is a requirement for our classification task, and necessary for our model to learn how to distinguish major-mergers from other types of galaxies. As there are many more galaxies in the simulation than just major-mergers, we use the number of major-mergers found in the \textbf{MM} sample selection as a guideline to define a control sample of non-interacting galaxies.

\begin{figure}
    \centering
    \includegraphics[width=0.5\textwidth]{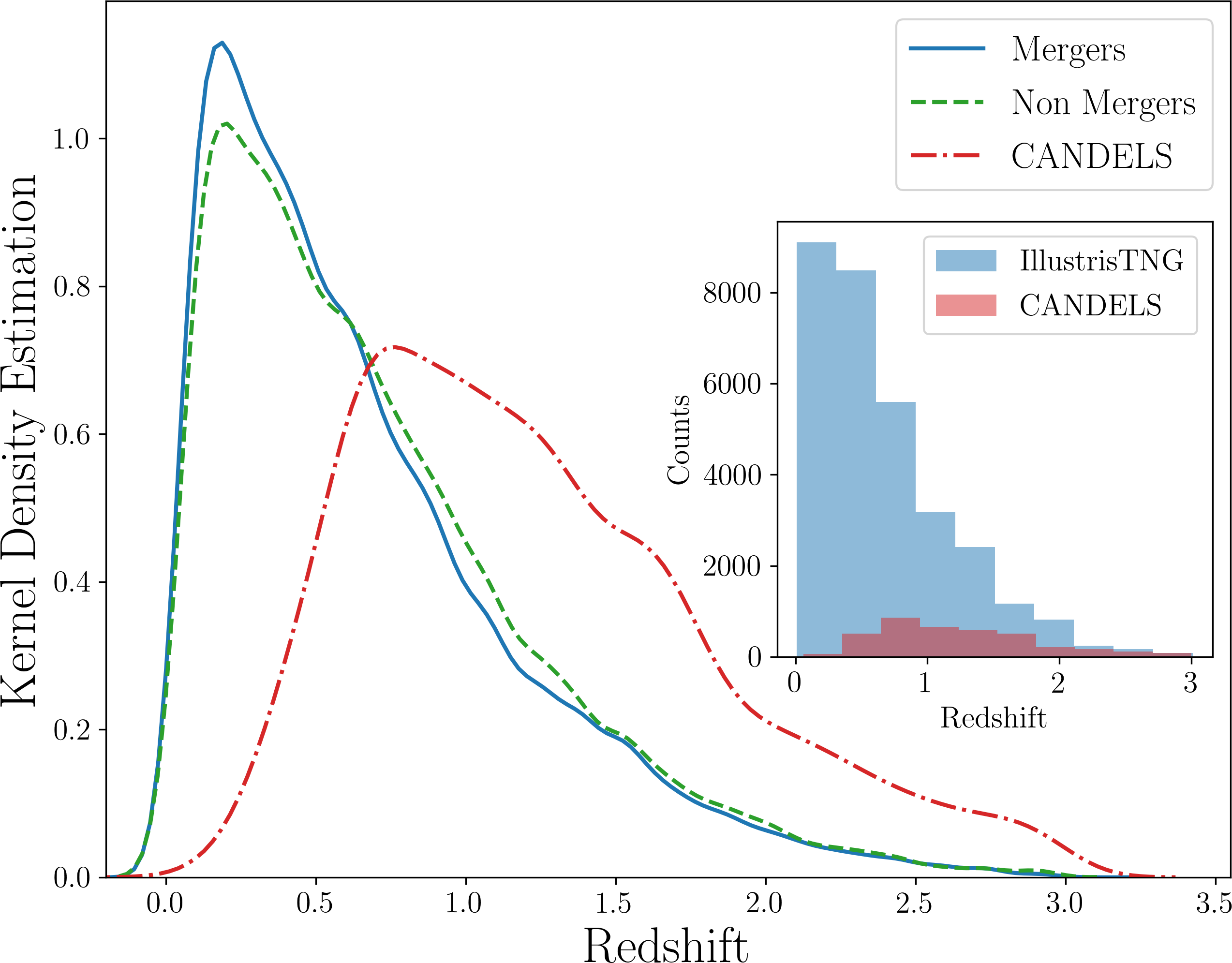}
    \caption{Redshift distribution for the simulated Major Merger sample (blue solid line), simulated non-interacting sample (green dashed line) and the CANDELS sample (red dot dashed line). The redshift distribution for our IllustrisTNG mergers and non-merger samples are by construction very similar. We also display the CANDELS redshift distribution to show that it does not match the redshift distribution of the samples used for training, but its numbers are within the range of the simulation distribution, as demonstrated by the unnormalized redshift histogram in the inner plot, showing all the IllustrisTNG galaxies in blue and CANDELS galaxies in red.}
    \label{fig:redshiftdist}
\end{figure}

First we apply redshift and stellar mass cuts to select galaxies in the same range as the  \textbf{MM} sample, with $z < 3$ and $M_* \ge 10^{10} \ \rm M_\odot$. Next, we clean this pre-selection from interacting galaxies as best as possible. This can not be done by just simply removing the galaxies found in the  \textbf{MM} sample from this new selection as there are other mergers occurring, with lower mass ratios, and cases where a merger event can have longer timescales than  $\tau_{\rm obs} \pm 0.3 \ \rm Gyr$, for selecting the  \textbf{MM} sample. This means that it is possible to have merging morphologies with broader timescales in the simulation. Thus, to solve this we do a broader search of merging galaxies, looking at all mass ratios and mergers occurring in $\pm 0.5 \rm \ Gyr$. Then, we proceed to remove all galaxies found in this way from the initial redshift and stellar mass cut. The resulting sample is then separated in the same bins of redshift as the major merger sample, enabling us to draw randomly the same number of galaxies for each redshift bin in order to construct a sample that has a similar redshift distribution, as shown in Fig (\ref{fig:redshiftdist}) (in the outer plot by the blue solid line and green dashed line, for mergers and non-mergers, respectively). 

Nevertheless, these selections are made only within the simulation merger trees. We still need to produce the imaging data that will be used to train our model. However, it is important first to define the data in which we are going to apply our model to make predictions, as we have to apply similar instrumental and observational effects in order to mimic the data the best way possible. In our case, we want to apply our model to galaxies in the CANDELS fields.

\subsection{CANDELS Fields}\label{sec:CANDELSFIELDS}

One goal of this work is to do predictions on CANDELS WFC3/IR imaging data \citep{Grogin2011, Koekemoer2011}. This consists of wide field data with enough depth to detect galaxies in the limit of our selection on the simulation data. This data was already used extensively within galaxy merger studies, with merger rates estimated up to $z \sim 6$ \citep{Duncan2019}. There are also visual morphology classification catalogues \citep{Kartaltepe2015}, photometric redshifts and stellar mass estimates \citep{Duncan2019}, which are essential if we want to make the same selection cuts as the ones done in IllustrisTNG simulation data, as we are only interested in predictions on a similar parameter space. 

Here our selection is similar to the one applied to the IllustrisTNG merger trees, with the exception that we do not use any merger classifications available to select it. The first step consists in removing all objects that have problems with quality flags in the original photometry catalogue and the \cite{Kartaltepe2015} catalogues, as we want to avoid edges, artifacts and stars. Then, we apply a magnitude cut in the H band of $H < 24.5 \rm \ mag$ following the same cut used in \cite{Huertas-Company2016} and \cite{Kartaltepe2015}. A signal-to-noise (SNR) cut of $\rm SNR > 20$ is also applied, as the magnitude cut would bias the SNR of our sample against extended sources. Then we proceed with the same cuts we made to the IllustrisTNG selection, using $z < 3$ and $M_* > 10^{10} M_\odot$. {\color{black}{This results in a sample of 3759 galaxies wish high enough SNR.}}

Fig. (\ref{fig:redshiftdist}) shows the redshift distribution of this subsample of CANDELS galaxies (red dot dashed line). It can be seen that this redshift distribution does not match the redshift distribution for IllustrisTNG galaxies. However, the inner plot shows an unnormalized redshift histogram of IllustrisTNG (blue) and CANDELS galaxies (red), which demonstrates that our training sample of IllustrisTNG galaxies is large enough to have at least similar galaxy counts to the CANDELS sample at higher redshifts. One might argue that it would be ideal to construct the training sample with the same redshift distribution as the data we are planning to do predictions with, but in this case, we are limited by resolution, which requires us to limit the scope to massive galaxies ($M_* \ge 10^{10} M_\odot$) only. At the same time, we are not introducing redshift information during training, apart from embedded instrumental and cosmological effects, so the variability on merger morphologies available in the regime where both redshift distributions disagree ($z < 0.5$) is essential to the learning model.

In the training step we tested matching the redshift distribution of the training sample with the CANDELS redshift distribution by removing low redshift galaxies from the training sample. However, our findings suggest that the performance of the model suffers from the smaller training sample by over predicting mergers at low redshifts. This is due the lack of generalization by the model when limited to smaller training samples. In this way, additional tests with different training samples are left for future work. Even though these galaxies can be considered intrinsically different, their morphologies are degenerate.

Finally, we produce cutouts from the imaging data that represents a field of view of $50 \ \rm kpc \ \times 50 \rm \ kpc$ using available redshift. {\color{black} In this way, we choose to rely on the redshift information available instead of using any assumption about the sizes of galaxies in our samples, as it is difficult to define it when two or more galaxies are interacting in the field of view. By using this approach, we are also preserving relative sizes between galaxies within our samples, which might provide important information for the network to use during the classification}. As we are using CANDELS Near IR data, we proceed to produce galaxy images from IllustrisTNG and apply instrumental and cosmological effects to the images so that they are a realistic representation of CANDELS galaxies.

\subsection{IllustrisTNG Imaging Data}\label{sec:imaging_data}

We take advantage of the tools available in the IllustrisTNG API and website to select stellar maps for a given object in the simulation. The 'Galaxy and Halos Vizualization'\footnote{\url{http://www.tng-project.org/data/vis/}} \citep{Nelson2018} tool enables us to select a galaxy by combining the simulation run, snapshot and subfind identification to visualize a given object in several filters. It uses a pipeline coupled with CLOUDY \citep{Ferland2017} photoionization code and Flexible Stellar Population Synthesis (FSPS)\footnote{
FSPS uses Kroupa IMF whilst stellar masses in our CANDELS catalogs are measured with Chabrier IMF, a $\sim 5\%$ offset is expected.
} through \texttt{python-fsps} \citep{Conroy2009, Conroy2010}, a stellar population synthesis code, generating stellar density maps for the appropriate ages and metalicities (in rest or observational frames), as selected by the chosen filter, refer to \cite{Nelson2018} for details.  However, this procedure has its limitations, as described earlier, as it does not include a full radiative transfer treatment, and does not account for dust.

%%(Model A in \cite{Nelson2018a}). 

This could impact some of the morphologies presented, especially for the star forming galaxies. Although studies using IllustrisTNG mocks generally use a complete radiative transfer approach for galaxies with high star formation rates \citep{Nelson2018a, Rodriguez-Gomez2018, Huertas-Company2019}, we limit our sample only to near-infrared filters as a way to mitigate potential biases due the absence of dust in our treatment. Thus, \cite{Bottrell2019} {\color{black} shows that realistic instrumental effects, such as noise and an appropriate PSF}, are more important than radiative transfer effects when training deep learning models, where the slight improvement in performance comes with a huge computational cost of producing galaxy mocks with full radiative transfer, especially for  large samples of galaxies. Moreover, we do not explicitly use any color information in our model. In this way, one might use our galaxy mocks as stellar density maps, which will be closely related to the true morphology of the galaxy.

The following is a brief overview of our complete mock pipeline. The first step consists of the selection pipelines described in \S \ref{subsec:major_mergers} and \S \ref{subsec:non_interacting}. The result of the selection is a list with each galaxy snapshot, subfindID and redshift. This is then fed to the Illustris API, requesting the mock produced by the Galaxy and Halos Vizualization pipeline. These images have field of views of $120 \ \rm kpc \ \times 120 \ \rm kpc$ and are imaged in the observed frame for the HST F125W and F160W filters, which are available for the CANDELS fields. For each subsample, we randomly request 80\% of the galaxies as face-on and 20\% as edge-on, as we do not have the freedom to choose arbitrary orientations using this tool\footnote{As this paper goes to press a new feature in IllustrisTNG API enable the user to use different projections and orientations instead of only face-on and edge-on orientation. This was not available when we generated our sample and we advise anyone doing a similar approach to use this new feature instead of only edge-on and face-on cases.}. {\color{black}{This proportion of face-on and edge-on galaxies is draw from axis ratio statistics from real galaxies in the CANDELS fields \citep[e.g.,][]{Ravindranath2004, Mowla2019}}}. This produces a set of clean images from the IllustrisTNG in the appropriate band, with cosmological dimming and k-correction applied. However, it is necessary to apply transformations in order to make mocks of these images as if they were observed by HST.

We apply cosmological geometric effects based on 'redshifting' \citep[e.g.,][]{Conselice2003, Barden2008} approaches and add features of image realism \citep{Bottrell2019} by appropriately simulating characteristics of CANDELS images, such as noise, PSF and adding the resulting image to a patch of the sky from the CANDELS fields. First, for each galaxy we apply a random rotation to the image following a crop to $50 \ \rm kpc \  \times 50 \rm \ kpc$ field of view for both filters. The reason why images have such large fields of view is to have an adequate window for image transformations. If one would crop a galaxy image after a random rotation, artifacts would be noticeable around the edges, especially for cases with intermediate rotation angles. Then, as we know the exact pixel scale of the clean image, we can transform it to $60 \ \rm mas/pixel$ HST WFC3/IR pixel scale and apply PSF effects by convolving it with a simulated PSF produced with \texttt{TinyTim}  \citep{Krist2004}. 

%\begin{figure}
%    \centering
%    \includegraphics[width=0.5\textwidth]{fluxogram.pdf}
%    \caption{Complete mock pipeline with all steps from the selection of the sample to the packaging and preparation for the CNN. Here clean images from the Illustris API are post-processed to include instrumental and redshift effects to mimic CANDELS data in each of the bands used. This includes rotation, pixelscale transformation and noise. The result of the pipeline is a package that is ready to be used by our CNN training pipeline.}
%    \label{fig:fluxmocks}
%\end{figure}

Noise is then added by converting the image to $e / s^{-1}$, multiplying it by an appropriated exposure time, and drawing a sample of it from a Poisson distribution. This is done to ensure that our mock images have similar shot noise to the real data. Then the resulting distribution is added to a empty sky region of the CANDELS fields. This region is selected randomly from a pool of pre-prepared regions. This is necessary, as the CANDELS fields are produced by a stack of multi-epoch sky subtracted images, which creates correlated noise \citep{Koekemoer2011}. {\color{black}{These regions are empty since we expect the impact from crowding to be small in the redshift range probed here. \cite{Bottrell2019} shows that the presence of neighbor sources during training is important for the success of the deep learning model, but their simulations are limited to low redshifts. However, we show in \S \ref{subsec:preds_on_illustris} that the presence of crowded sky regions impacts the model negatively.}}

%We measure a SegMap with GalClean and proceed to the final stage of the pipeline, where it prepare the image to be used by the CNN. 

After all of these effects are introduced to the image, we prepare it for the CNN by re-sampling it to 128x128 pixels. This is the same as changing the pixel scale once more, but in most cases we are oversampling the image, as by this stage all images should be smaller than 128x128 pixels, thus we are not losing information by doing this. This particular resolution is selected so as to provide the CNN with the possibility of having more convolutional layers. Then, we package the whole sample in a HDF5 file with its train, test and validation split, including normalization. This is the package that is then used by the CNN.

The result of the selection and imaging data pipeline is summarized in Table (\ref{tab:sample_sum}).

\begin{table*}[]
    \centering
    \begin{tabular}{|p{2.3cm}|p{1.7cm}|c|c|c|c|c|c|c|c|c|c|c|c|}
        \hline
             \multirow{2}{2cm}{\textbf{Redshift}} &  \multirow{2}{2cm}{\textbf{Snapshots}} & \multicolumn{3}{c|}{\textbf{Number of Galaxies}} &  \multicolumn{3}{c|}{\textbf{Before Merger}} & \multicolumn{3}{c|}{\textbf{After Merger}} &  \multicolumn{3}{c|}{\textbf{Non Interacting}} \\
             \cline{3-14} 
             & & Train & Test & Val & Train & Test & Val & Train & Test & Val & Train & Test & Val \\
             \hline
             $ 0.0 \le z < 0.5$ & 99-66  & 19633 & 4257 & 4214 & 5331 & 1076 & 1171 & 4966 & 1117 & 1035 & 9336 & 2064 & 2008 \\ \hline
             $ 0.5 \le z < 1.0$ & 67-51  & 13410 & 2837 & 2931 & 3240 & 669 & 726 & 3434 & 697 & 755 & 6736 & 1471 & 1450 \\ \hline
             $ 1.0 \le z < 1.5$ & 50-41  & 6127 & 1342 & 1299 & 1377 & 292 & 295 & 1599 & 348 & 320 & 3151 & 702 & 684 \\ \hline
             $ 1.5 \le z < 2.0$ & 40-33  & 2821 & 563 & 551 & 715 & 148 & 122 & 681 & 141 & 137 & 1425 & 274 & 292 \\ \hline
             $ 2.0 \le z < 2.58$ & 33-27 & 993 & 213 & 216 & 257 & 62 & 60 & 240 & 44 & 44 & 496 & 107 & 112 \\ \hline
             $ 2.58 \le z < 3.0$ & 28-25  & 210 & 44 & 45 & 57 & 12 & 14 & 51 & 7 & 9 & 102 & 25 & 22 \\ \hline

             \multicolumn{2}{|c|}{{\textbf{Totals}}} & 43194 & 9256 & 9256 & 10977 & 2259 & 2388 & 10971 & 2354 & 2300 & 21246 & 4643 & 4568 \\
             \cline{3-14}
              \multicolumn{2}{|c|}{{}} & \multicolumn{3}{|c|}{61706} & \multicolumn{3}{|c|}{15624} & \multicolumn{3}{|c|}{15625} & \multicolumn{3}{|c|}{30457} \\
             \hline
    \end{tabular}
    \caption{Summary of the IllustrisTNG samples of major-mergers and non interacting galaxies separated in redshift bins, label and the Training, Testing and Validation subsamples.}
    \label{tab:sample_sum}
\end{table*}{}

\section{Methods} \label{sec:methods}

We employ a Deep Learning approach with Convolutional Neural Networks (CNNs) to our images, a state of the art tool to solve computer vision problems \citep{Goodfellow-et-al-2016} that is gaining popularity among galaxy merger studies \citep{Ackermann2018, Pearson2019, Bottrell2019}. In a CNN, convolutional layers use convolution operations on multidimensional data, such as images, to extract features that can then be used for classification tasks in regular fully connected layers at the top of the CNN architecture. The convolutional part of the network can be divided into convolutional blocks, which can then nest more types of layers than just convolutional layers. However, each block is generally limited to probe a specific resolution range of the input data. Pooling operations are usually located between convolutional blocks with the goal of changing the input image to a lower (or higher) resolution. How these blocks and layers are organized and how wide the network is, including the number of filters, size of the kernels, and other properties, are defined by hyperparameters.

We briefly describe our method for finding a good model with an optimization approach in \S \ref{subsec:bayesian}, together with a short description of each hyperparameter; We describe the metrics used to evaluate the performance of our models and the architecture found by our optimization approach in \S \ref{subsec:bestmodel}.

\subsection{Bayesian Optmization of Hyperparameters}\label{subsec:bayesian}

Generally, CNNs and other Deep Learning methods are regarded as black boxes since their parameters are adjusted by an automated training process in order to maximize its performance, with little control over it apart from the architecture of the network. Its architecture is defined by a set of parameters that control how big a network is, how many layers there are, the learning rate and batch size, among other configurations. The results produced by a network model are highly dependent on its hyperparameters, so it is of utmost importance to fine-tune them as best as possible \citep{Hacohen2019}. Unfortunately, there is no method that is capable of finding the best set of hyperparameters without training the network and assessing its performance. Often, this is done by bruteforce methods such as grid searches, where a large domain of possible values for each hyperparameter is defined and portions of the domain are evaluated by training the corresponding network. If a high number of hyperparameters are present, the result is a very expensive task and might not lead to the best model.

To avoid this treatment, we use a Bayesian Optimization approach to find a good set of hyperparameters by  modeling our architecture as a surrogate gaussian function $g(\mathbf{x_1}, ..., \mathbf{x_n})$, where $\mathbf{x_1}, ..., \mathbf{x_n}$ are the hyperparameters. Each possible combination of hyperpameters is a different model. This function is very expensive to evaluate, but with few samples it is possible to reach a set of hyperparameters that best optimizes the performance of the model by updating the posterior at each sample, using it to make informed guesses for the next observation. This technique is faster and can yield a set of hyperparameters that results in models with better performances than ones optimized manually, reducing the number of configurations necessary to reach a good model  \citep{Larmarange2017}.

\subsubsection{Hyperparameters}

We first define what will be considered a hyperparameter in our architecture by defining what aspects of it can be changed, setting a domain for each case. Here we briefly describe each of the hyperparameters of the architecture while a summary is displayed in Table (\ref{tab:architecture}).

We define a convolutional block as a group of convolutional layers that probe similar input resolutions. Each block is separated by pooling layers that change the size of the input for the next block by a factor of 2. The number of convolutional blocks, \texttt{number\_conv\_blocks}, is one of the main hyperparameters to define how long the convolutional portion of the network will be. Thus, the number of layers in each block, \texttt{number\_conv\_per\_block} is also a hyperparameter. Every convolutional layer in a given block has the same number of filters and kernel size. The possible number of blocks varies between 1 and 5 while each block can have from 1 to 3 convolutional layers. Convolutional blocks not only group convolutional layers, but their activation and other auxiliary counterparts as well. Additionally, we set the number of filters in the first convolutional block, \texttt{initial\_number\_filters}, and the kernel size of the first convolutional block, \texttt{initial\_kernel\_size}, as hyperparameters. In a analogous way to the number convolutional layers, we consider the number of fully connected layers, \texttt{number\_fullyconnected\_layers}, and their size, \texttt{size\_fullyconeccted\_layers}, as hyperparameters as well. 

In neural networks, an optimizing function is used to maximize the performance of the network (minimize an error function). There are several distinct methods to accomplish this and different methods work better for different problems, as they represent strategies to find minima in the topology generated by parameters in parameter space. Here we choose from a pool of all optimizers available in Keras \citep{chollet2015keras} and let it also act as a hyperparameter of the architecture, even though it is not usually considered a hyperparameter.

We dedicate two hyperparameters to control the regularization of the architecture, namely the L2 regularization $\lambda$ term, \texttt{l2\_regularization}, and the dropout rate, \texttt{dropout}. The former act as a way to regularize the weights of the convolutional portion of the network by adding a penalty to the loss function in order to prevent spiked weights in favor of more diffuse configurations, while the later applies regularization to the fully connected layers by deactivating a percentage of the neurons for each layer equal to the dropout rate (\texttt{dropout}). By using dropout we will also be able to assess uncertainties in the network predictions. This is done by measuring probability distributions for each prediction by running the model for the same input with the dropout layers several times, as each time only portions of the fully connected layers are going to be used by the model. This approach is known as a Monte Carlo dropout \citep{Cook2000, Huertas-Company2019}.

Finally, we set a range of possible batch sizes, \texttt{batch\_size}, and possible initial learning rates, \texttt{initial\_learning\_rate}, as hyperparameters.

\begin{table}[]
    \centering
    \begin{tabular}{|r|c|}
        \hline
         Hyperparameter & Best Model  \\
         \hline
          batch\_size  & 256 \\
          number\_conv\_blocks  & 2 \\
          number\_conv\_per\_block  & 2 \\
          initial\_number\_filters  & 32 \\
          initial\_kernel\_size   & 11  \\
         
          number\_fullyconnected\_layers & 2 \\
          size\_fullyconnected\_layers & 1024 \\
          optimizer   & Adadelta \\
          initial\_learning\_rate  & 0.1 \\
          l2\_regularization  & 0.62 \\
          dropout & 0.38 \\
         \hline
         
    \end{tabular}
    \caption{Set of hyperparameters of our architecture and the best parameters found by doing Bayesian Optimization. }
    \label{tab:architecture}
\end{table}{}

\subsection{Performance Metrics and Best Model}\label{subsec:bestmodel}

In order to evaluate each of the possible models within our domain of hyperparameters, we first define how our models are going to be evaluated, since the Bayesian Optimization employed here runs as an automated process which tries to find the set of hyperparameters resulting in the best performance. This is assessed by training the network as a binary classifier of \textbf{MM}/\textbf{NM}  (see \S \ref{subsec:major_mergers} for definitions) with the training sample and performance evaluated in the testing sample. As we are not concerned with class imbalance problems at the moment, we simply try to minimize the loss function within our architecture. Models with low loss will represent models with high performance metrics. We also track the accuracy, precision and recall of each model, which inversely follow the loss very closely.

We perform the Bayesian optmization in the domain described with the GPyOpt python package \citep{gpyopt2016}. The model with the lowest validation loss is shown in Table \ref{tab:architecture}. 

\subsection{Bayesian Neural Networks}\label{sec:baynet}

Even though we carry out the hyperparameter optimization with the binary \textbf{MM}/\textbf{NM} classification, it is also important for us to probe if our CNN is capable of separating merger classes into further sub-classes, where galaxies are undergoing mergers at different stages. An easy distinction that we use from our selection procedure (Section \ref{subsec:major_mergers}) is to have a \textbf{BM}/\textbf{PM}/\textbf{NM} classifier. We follow a similar approach as is done by \cite{Huertas-Company2019}, where a hierarchy of binary classifiers are used to develop classifiers that are specialized in a specific separation task. In our case, this means that we will have a \textbf{MM}/\textbf{NM} classifier trained with all our sample and another one trained only with mergers to separate them into \textbf{BM}/\textbf{PM}. Then, the output for this set of binary classifiers can be combined with Bayes Theorem to yield the probability in each merger class by:

\begin{equation}
    P(\rm \mathbf{BM}) = P( \mathbf{MM}) \times P \left ( \frac{\rm \mathbf{BM}}{\rm \mathbf{MM}} \right), 
\end{equation}

\begin{equation}
    P(\rm \mathbf{PM}) = P( \mathbf{MM}) \times P \left ( \frac{\rm \mathbf{PM}}{\rm \mathbf{MM}} \right), 
\end{equation}

\noindent where the probability of being a \textbf{NM} is simply the output for the \textbf{NM} class in the \textbf{MM}/\textbf{NM} classifier. In this sense, the \textbf{MM} acts as a prior probability.

By combining multiple binary classifiers together to do multi-class classification we are combining models refined to perform very specific tasks instead of using only one classifier that has to share all its weights and parameters among all classes. However, even though in some cases the output probabilities will not have any meaning, they can still be used to investigate the classification process. For example, a relatively high $P(\mathbf{PM})$ value for \textbf{NM} galaxies might indicate that their morphology has aspects resembling a disturbed galaxy. A high value of $P(\mathbf{BM})$ in a \textbf{NM} galaxy might indicate that the galaxy has companions. Nevertheless, this should not be common within the simulation data but might be useful when performing predictions in real data where no labels are available.

\section{Results} \label{sec:results}

With the architecture and the sample from the simulation described in Section (\ref{sec:illustris_data}), we train our model and explore how it performs in the validation sample. In this way it is possible to analyze how the model generalizes to simulation data it has not seen. This is necessary before we apply it to real data. After checking if the results are what we would expect within the simulation, we apply our model to the sub-sample of galaxies from all the CANDELS fields as described in \S \ref{sec:CANDELSFIELDS}. 

\subsection{Predictions using IllustrisTNG}
\label{subsec:preds_on_illustris}

\begin{figure}
    \centering
    \includegraphics[width=0.5\textwidth]{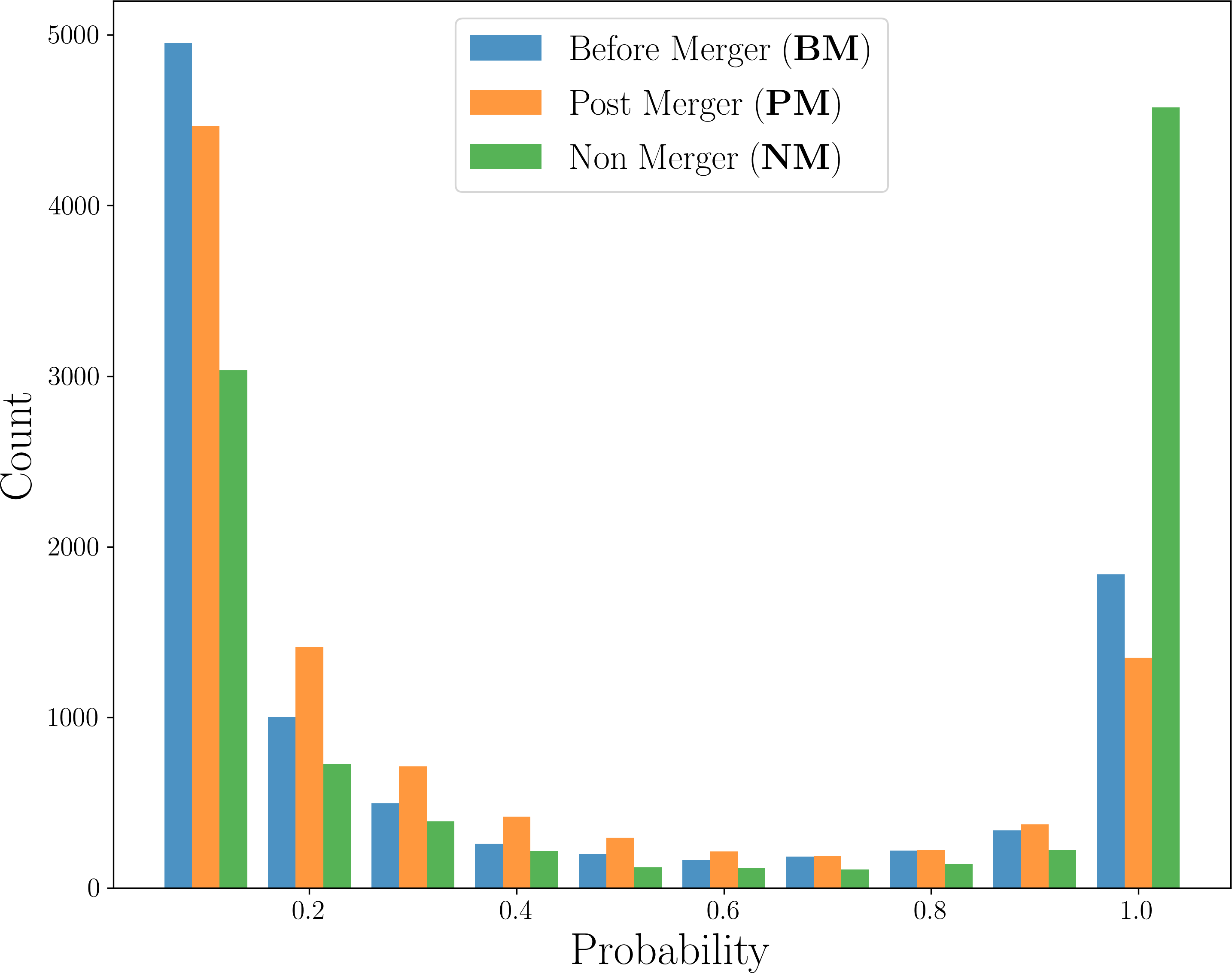}
    \caption{Class probability distribution of IllustrisTNG galaxies in the validation sample for each class in bins of 0.1 probability. This shows that our network has high confidence in the \textbf{NM} classifications whilst the probability distribution for the merger classes are more spread out. There is also a discrepancy between \textbf{BM} and \textbf{PM} in P $>$ 0.9, a sign that the \textbf{PM} class is the case that the network is less sure about, which has more ambiguity among the other types.}
    \label{fig:illustris_prob}
\end{figure}

By exploring how our models perform in the validation data, it is possible to identify its performance in a sample of galaxies from the simulation that the model has not seen during training or testing. Even though it should follow the performance of the testing set, this procedure enables us to verify if there are any biases in our set of classifiers. These, if present, can then be used to adjust predictions on real data later. We apply our model to the validation data to classify all galaxies in the sample in three classes: \textbf{BM}, \textbf{PM} and \textbf{NM}, as defined in \S \ref{subsec:major_mergers}. In Fig. (\ref{fig:illustris_prob}) we show the distribution of probabilities assigned to each class using predictions within our hierarchy of models, as described in \S \ref{sec:baynet}. {\color{black} We can see that the classifier is fairly balanced between \textbf{MM} and \textbf{NM}, which is expected since the distribution of our simulation data is balanced. However, when comparing merger sub-classes, the distribution is skewed towards \textbf{BM}, as the network is less sure about \textbf{PM} classifications. }

{}

\begin{figure}
    \centering
    \includegraphics[width=0.5\textwidth]{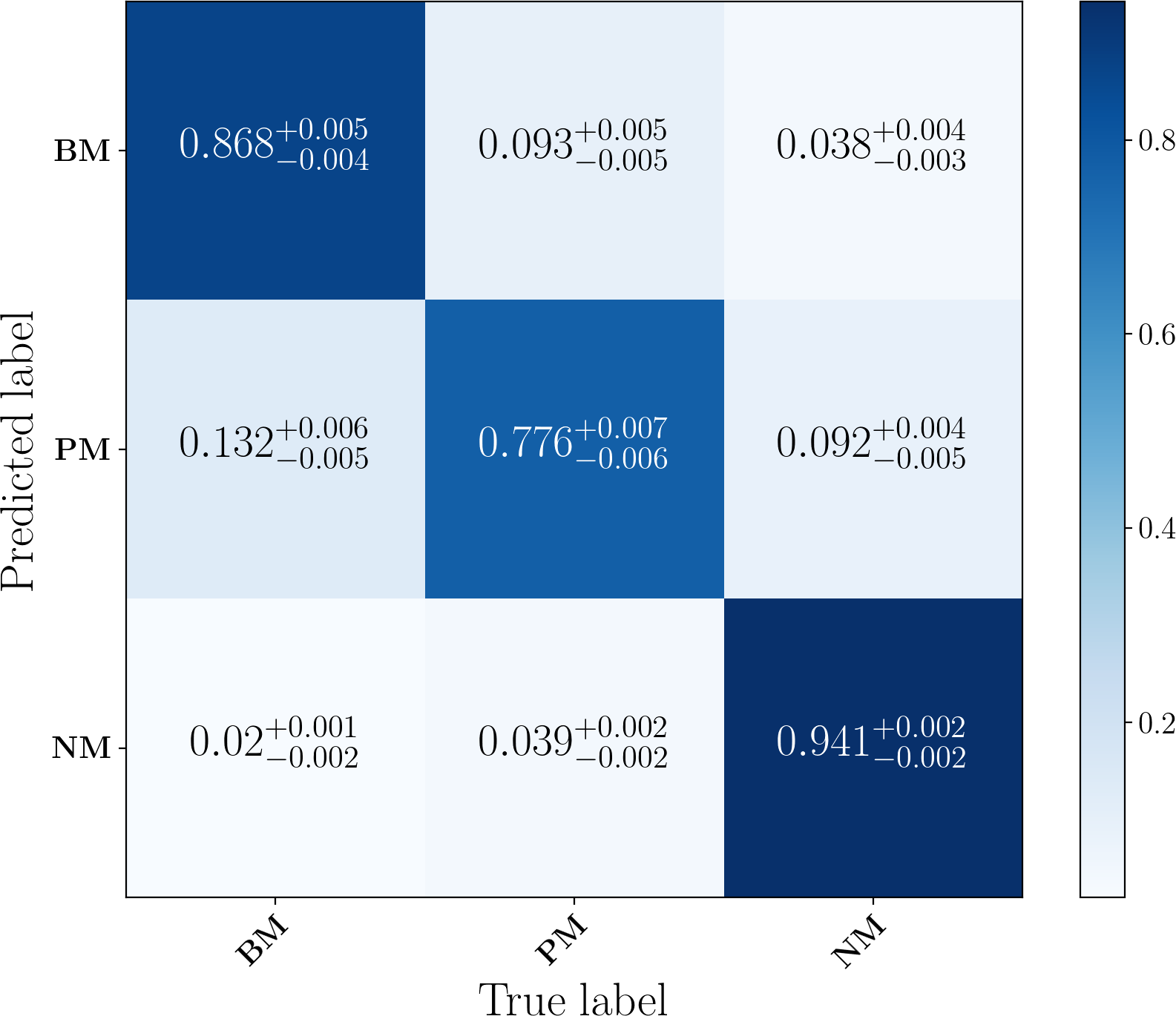}
    \caption{The normalized confusion matrix for our classifier hierarchy. Each column represents the true labels for each class while rows represent the predicted class. The diagonal of a multi-class classifier present the precision for each class, while other cells show the contamination between each possible pair of classes. It is important to note that almost two thirds of the contamination of \textbf{PM} happens with \textbf{PM} being classified as \textbf{BM}, which is still a merger classification. Errors shown are measured with the Monte Carlo dropout. This confusion matrix is measured within our balanced validation sample and do not represent the performance of the method with real galaxies. }
    \label{fig:cm_illustris}
\end{figure}

 The class probability distributions shown in Fig. (\ref{fig:illustris_prob}) are not enough to draw conclusions about our CNN's performance, we further explore performance metrics with our validation sample. We evaluate our hierarchy of models by looking at its normalized confusion matrix, which is shown  in Fig. (\ref{fig:cm_illustris}). The confusion matrix gives us an overview of the performance of the model by comparing the predicted labels with the true labels for each class. It shows this by listing the precision of each class in the diagonal, the fraction of correct classifications among all examples for the given class, while also showing the relative miss-classifications between each pair of classes. Our model is capable of identifying \textbf{BM} and \textbf{NM} types with 87\% and 94\% accuracy, respectively, with a contamination between both classes of less than 5\%. However, in the \textbf{PM} case, the model has a lower performance, with 78\% correct classifications with 13\%  contamination with \textbf{BM} and 9\% contamination with \textbf{NM}. Even though it has almost a 10\% performance difference with the other classes, almost two thirds of its miss-classifications are still merger classifications. Also, as in some cases the morphology of \textbf{PM} systems have no clear distortions, we therefore expected it to have some degeneracy with \textbf{NM} galaxies, while this is not true for the \textbf{BM} and \textbf{NM} classes.

%As a whole, the model's performance on the validation sample -- galaxies it never saw during training -- has higher precision than previous attempts. 

{}

\begin{figure*}
    \centering
    \includegraphics[width=\textwidth]{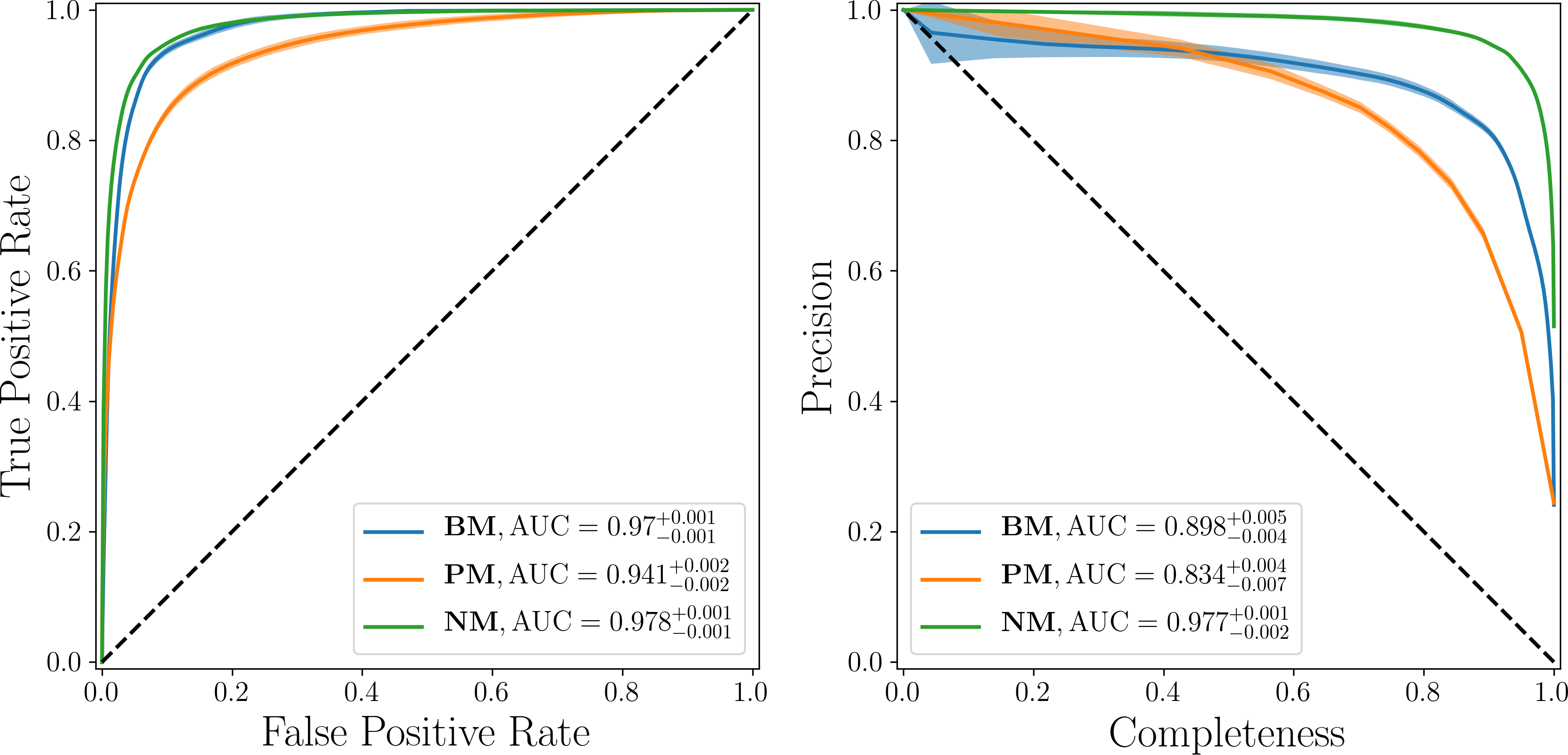}
    \caption{Performance metrics for classifications using the validation data. ROC curves for each class are shown in the left plot with \textbf{BMs}, \textbf{PMs}, \textbf{NMs} in blue, orange and green, respectively. The compromise between completeness and precision is shown in the right with the same color code. The performance shown here is based on the balanced validation sample, real galaxy samples will have very unbalanced configurations and hence this metric does not translate directly to applications on real galaxies.}
    \label{fig:illustris_metrics}
\end{figure*}

It is also useful to verify the model with other metrics, especially the Receiver Operating Characteristic curves (ROC curves) and Precision-Recall diagrams \citep{Powers2011}. These are important because they also take classification threshold into account, while the confusion matrix only uses one threshold specified before-hand (i.e predictions should be in binary form). In Fig. (\ref{fig:illustris_metrics}) we show ROC curves for each class in the left panel and the Precision-Recall curves in the right panel. Precision-Recall curves can also be though as Purity-Completeness diagrams, which are a more common convention in astronomy. As we are using Monte Carlo dropout, we have ways of estimating the uncertainty of our classifications. Due to this feature of our model, we can plot the mean curves for each diagram with confidence intervals. This can be seen in each of the plots in Fig. (\ref{fig:illustris_metrics}) by the shaded area, which represents $\pm 4 \ \sigma$ from the mean of the model, shown as a solid line. For the ROC curves, this uncertainty is very small and all classes follow a similar trend to what we might expect for a model with a confusion matrix equal to the one presented in Fig. (\ref{fig:cm_illustris}). The area under the curve is also shown in the legend. 

{}

For the Precision-Recall diagram in the right panel of Fig. (\ref{fig:illustris_metrics}), it is possible to check that the uncertainties in our model are more apparent in the region of high precision. This is due to the fact that in this regime the threshold is very high, limiting the model to only very precise classifications. This results in smaller sets of classified galaxies, with very poor completeness, that are more prone to variability. 

\begin{figure*}
    \centering
    \includegraphics[width=\textwidth]{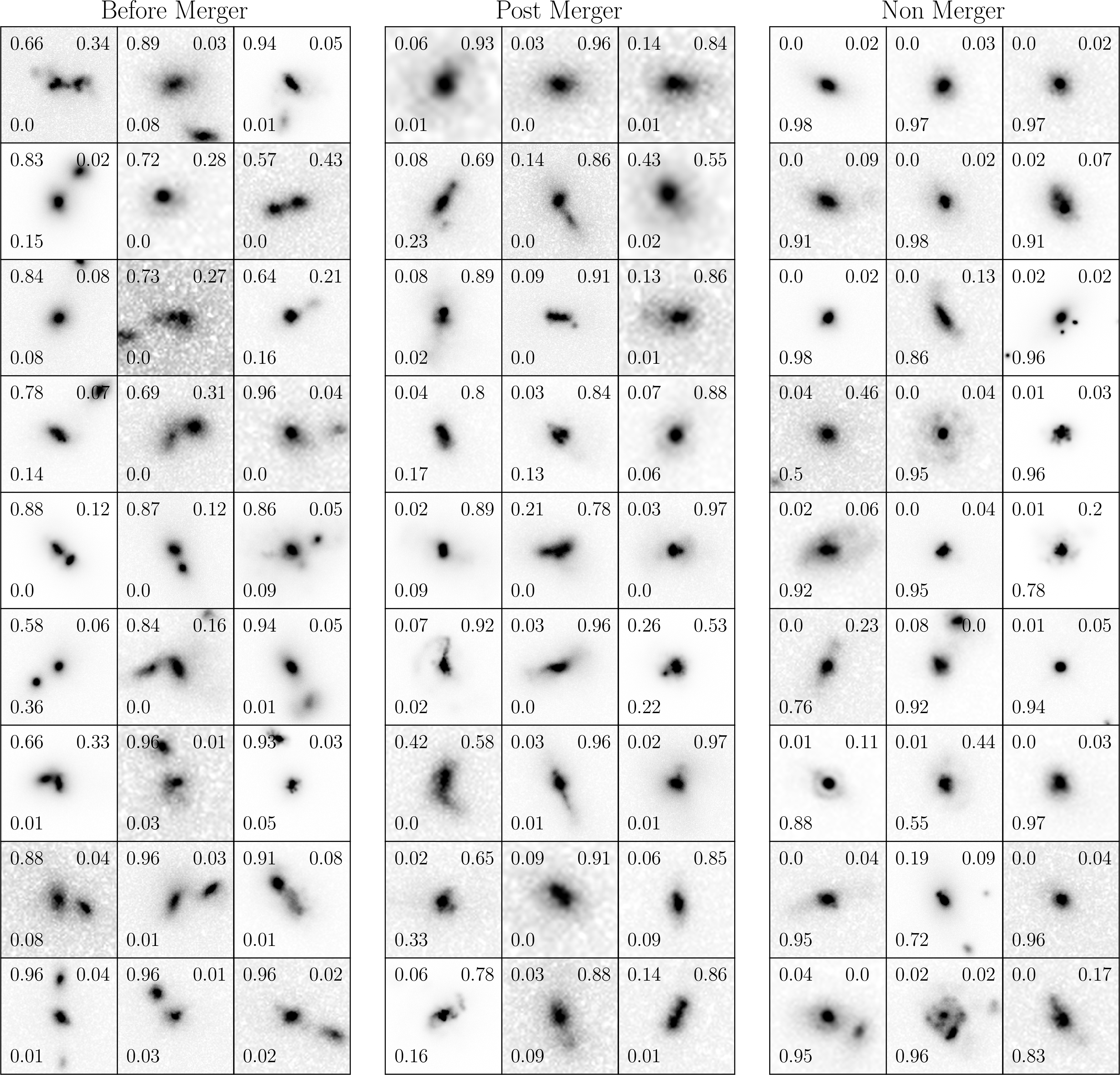}
    \caption{Mosaics for each class as classified by our model using simulated IllustrisTNG data. All galaxies were randomly drawn from the validation sample. In each galaxy image, all three probabilities are shown on each image. $\rm P( Before \ Merger) $, $\rm P( Post \ Merger)$, $\rm P( Non \ Merger)$, top-left, top-right and bottom, respectively. Varying signal-to-noise in the images are due to the varying intrinsic luminosity of the simulated galaxies or due to cosmological dimming.}
    \label{fig:illustris_mosaic}
\end{figure*}

For visualization purposes, we plot a mosaic of images with galaxies randomly drawn for each class in Fig (\ref{fig:illustris_mosaic}). Every galaxy plot shows the probabilities for the three classes, $\rm P( \mathbf{BM}) $, $\rm P(  \mathbf{PM})$, $\rm P(\mathbf{NM})$. Thus, as these galaxies are randomly selected, we also have cases that are miss-classifications. It is important to note that the threshold used here is the binary threshold, for probabilities $P > 0.5$, so this show the standard performance of the model, based on the confusion matrix of Fig. (\ref{fig:cm_illustris}).

%If one increases the threshold, the number of miss-classification would decrease while the size of the sample would also decrease. 

{}

\begin{figure*}
    \centering
    \includegraphics[width=\textwidth]{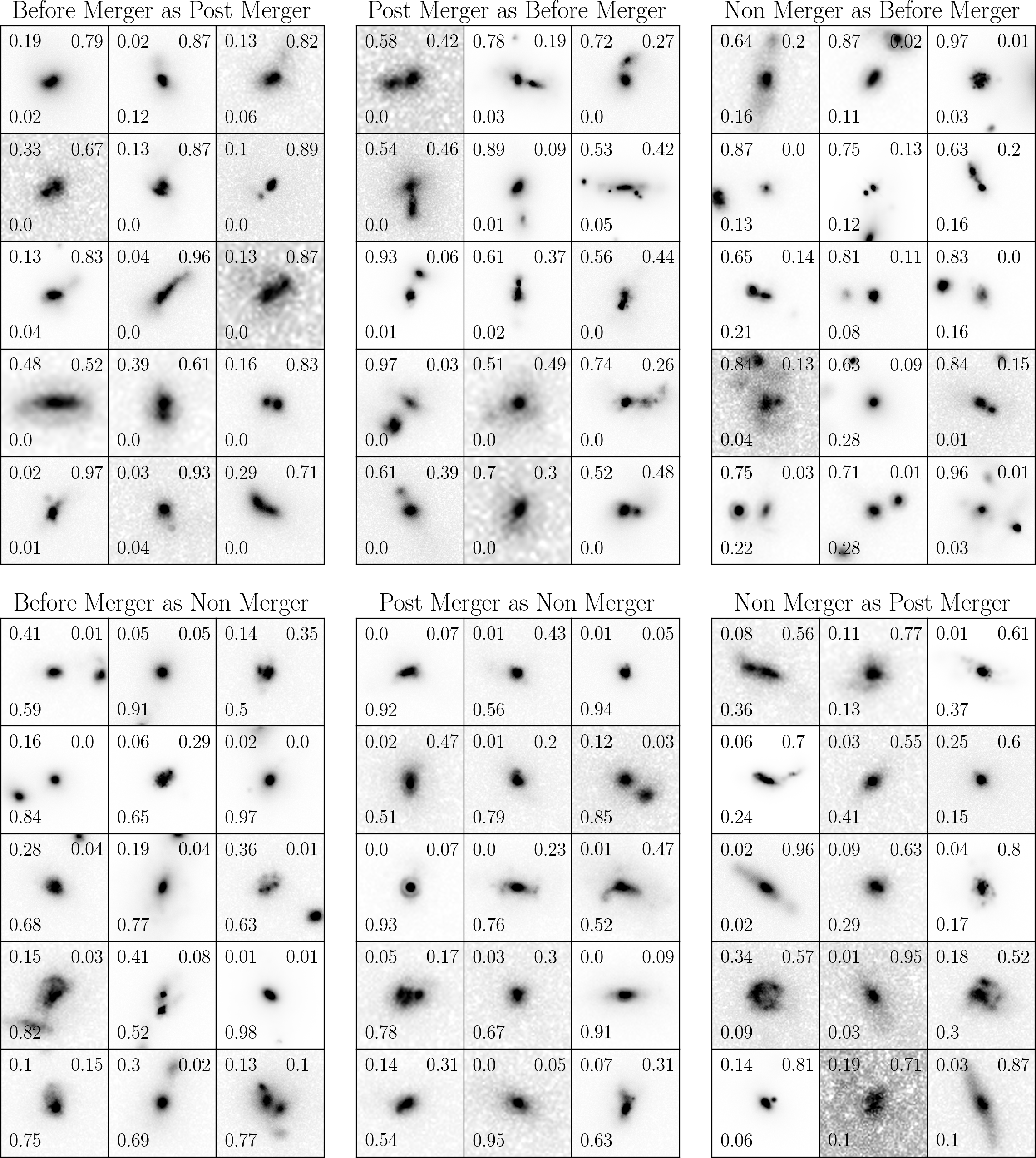}
    \caption{Mosaics for each possible case of miss-classification in the simulated IllustrisTNG data. Each title describes what is the truth class being miss-classified as a different class (truth class as wrong class) on given panel. All galaxies were randomly drawn from the validation sample for each specific case. In each galaxy image, all three probabilities are shown in each plot.  $\rm P( BM) $, $\rm P( PM)$, $\rm P( NM )$, top-left, top-right and bottom, respectively. Varying signal-to-noise in the images are due to the varying intrinsic luminosity of the simulated galaxies or due to cosmological dimming.  }
    \label{fig:illustris_mosaic_missclass}
\end{figure*}

It is also useful to characterize each type of miss-classification produced by the network. In our case, this represents 6 different kinds of miss-classifications, one for each possible pair of classes in our three class hierarchy. We plot in Fig. (\ref{fig:illustris_mosaic_missclass}) a panel of 15 miss-classified galaxies for each possible pair. The title of each panel refers to the true class, and what was the classification based on the probability from the model. Here, we see that the classifier uses very clear characteristics of merging for classifying galaxies as \textbf{BM}, as all galaxies misclassified as \textbf{BM} look as though they have two nuclei, or featuring two or more galaxies very close together. This even appear for \textbf{NM} systems classified as \textbf{BM}, a clue that our selection process for \textbf{NM} has some, even though small, contamination from galaxies with close companions. It is possible that the selection is not accounting for some types of mergers. Likewise, galaxies misclassified as \textbf{NM} are in general more symmetric than their true counterparts. For instance, \textbf{BM}s classified as \textbf{NM} still show companions and some sort of interaction, but are more symmetric than most \textbf{BM} in Fig. (\ref{fig:illustris_mosaic}).

We also see that \textbf{BM} systems classified as \textbf{PMs} show clearly signs of two nuclei, but for those which are closer together than regular \textbf{BM} systems. This is a sign of some degeneracy on the Sublink algorithm. Even if two galaxies are roughly in the same space, such that can still be regarded as two distinct galaxies. A similar pattern is seen in the case of \textbf{NMs} classified as \textbf{PMs}, as these non-interacting galaxies are more disturbed than their true counterparts. This shows us, overall, that the miss-classifications say a lot about how our model classifies a galaxy, as it follows properties that would also be used in visual classifications. Often, miss-classifications happen for cases where the morphology is really degenerate between classes, which would be expected. These are generally regarded as hard cases to learn, a natural limitation to the method based on visual structure, as they represent less than 3\% of the training data which is not enough to represent significant shift in the weights of the model.

{}

\begin{figure}
    \centering
    \includegraphics[width=0.5\textwidth]{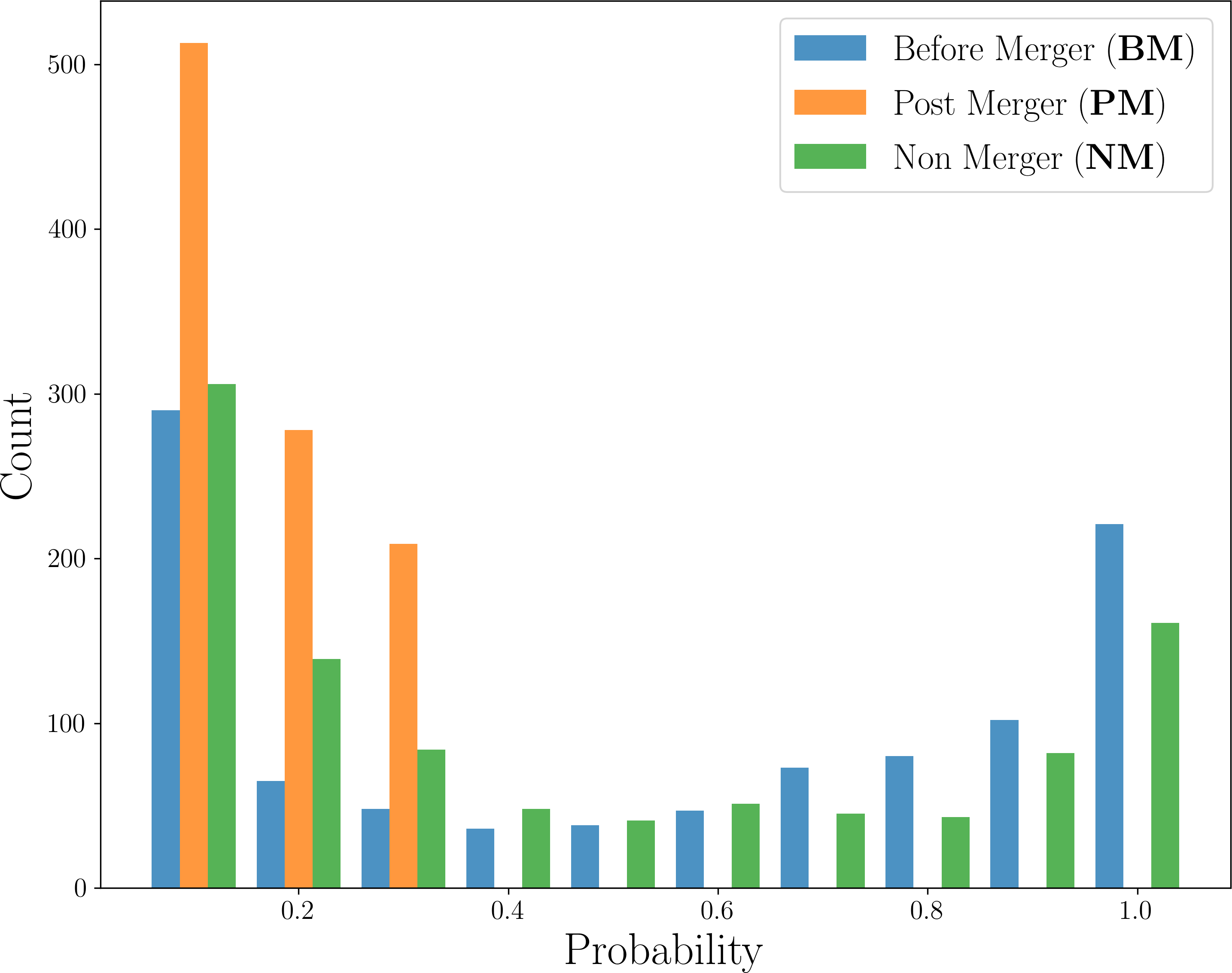}
    \caption{Mean Posterior Probabilities for all images in the random noise sample. Our hierarchy of model tends to classify most of the random noise images as \textbf{BM} and \textbf{NM} while none of the high probability noise images are classified as \textbf{PM}.}
    \label{fig:random_prob}
\end{figure}

Yet another meaningful test is to generate images of pure random noise to check how our methods deal with images that are not representative of the parameter space we are interested in. As the model has to assign probabilities that sum to 1 to any image given to it, it will by design likely classify a random noise image as one of the possible classes. By generating a relatively large sample of random noise images we can inspect the output probabilities to check the behavior of the network in this case. To do so we generate 1000 random images within two filters each\footnote{We also investigated completely random noise and different images for each filter and the same random noise for both filters, with similar results.}, representative of the filters of our regular input data, and feed it to the network. We explore the probability distribution of each class in Fig. (\ref{fig:random_prob}).

{}

These probabilities show that our model tends to classify $\sim 60\%$ of the noisy images as \textbf{BM} and $\sim 40\%$ as \textbf{NM}. This is a good sign, as we have two opposite classes that show a similar behavior towards noise. The network did not classify any of the input random images as \textbf{PM}, where the maximum probability among all classifications was $P(PM) = 0.48$. This means that we can be fairly secure that miss-classification of \textbf{PM}s due to image quality effects, like noise, will be rare.

%This information is specially useful when we, in the following section, will use one of the merger classes to measure merger rates. 

\begin{figure}
    \centering
    \includegraphics[width=0.5\textwidth]{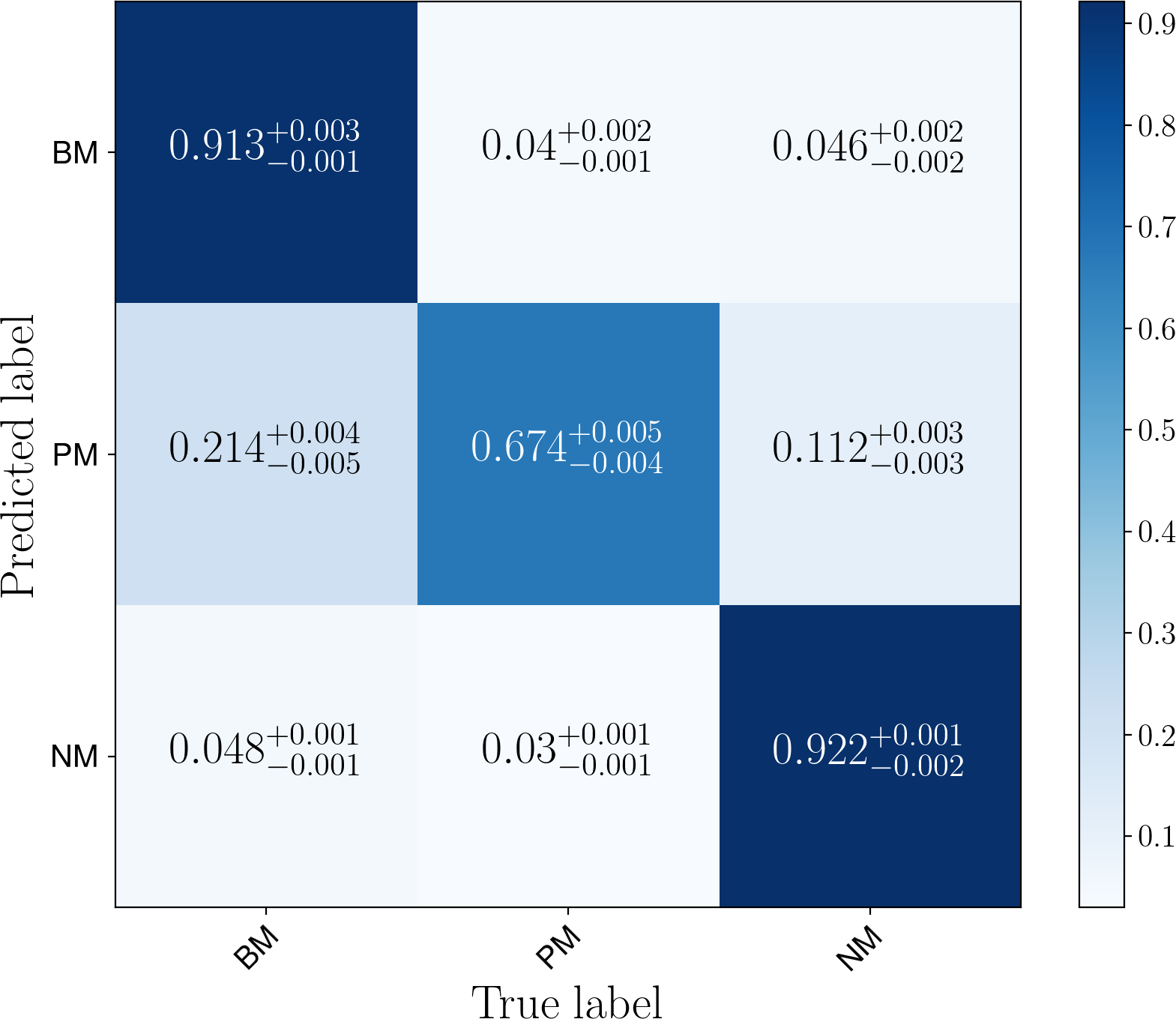}
    \caption{The normalized confusion matrix for our classifier hierarchy trained with simulated galaxies included in crowded patches of the sky from the CANDELS fields. Each column represents the true labels for each class while rows represent the predicted class. The diagonal of a multi-class classifier present the precision for each class, while other cells show the contamination between each possible pair of classes. It is important to note that almost two thirds of the contamination of \textbf{PM} happens with \textbf{PM} being classified as \textbf{BM}, which is still a merger classification. Errors shown are measured with Monte Carlo dropout. }
    \label{fig:crowded}
\end{figure}

{\color{black}{
Finally, we assess how the presence of crowded sky regions impacts our model classification. \cite{Bottrell2019} shows that the presence of contamination from neighboring sources is important during training when using simulated galaxies at low redshift. To show if this statement is true for the data used here, we retrain our model with a new dataset of simulated galaxies prepared with random patches of the sky from the CANDELS fields. These random regions are selected by searching for places that are centrally empty but have neighbor sources around the center. 

The confusion matrix displayed in Fig. (\ref{fig:crowded}) shows that in this situation the classification precision of \textbf{BMs} slightly improves from 87\% to 91\%, whilst \textbf{PMs} and \textbf{NMs} decrease, from 78\% to 67\% and 94\% to 92\%, respectively. Even though our results for the presence of crowded backgrounds diverge from what is shown in \cite{Bottrell2019}, we attribute it to the difference in scope of our data. We probe higher redshifts ($0 < z \le 3$) and different wavelengths with simulated galaxies from cosmological simulations, which have lower resolution than galaxy-galaxy simulations. This experiment, however, shows that in crowded regions we should expect our model to display worse performances for \textbf{PMs}. In the case of galaxies in the CANDELS fields, we are selecting small field of views and expect low contamination from crowded regions. As the overall results are worse with crowded regions of the sky, we conduct the rest of the paper with the class hierarchy trained with the original dataset. 
}}

{}

It is important to note, however, that all performance metrics shown in this section are valid within the scope of our simulation validation sample. This needs to be taken into account when applying our classifier hierarchy to real data, as we expect to have an unbalanced sample of BMs, PMs and NMs. As we do not have ways to directly assess the performance of this classifier in the real data, we have to make comparisons with visual classifications and galaxy merger rates to test it.

\subsection{Predictions on CANDELS}\label{subsec:preds_CANDELS}

\begin{figure}
    \centering
    \includegraphics[width=0.5\textwidth]{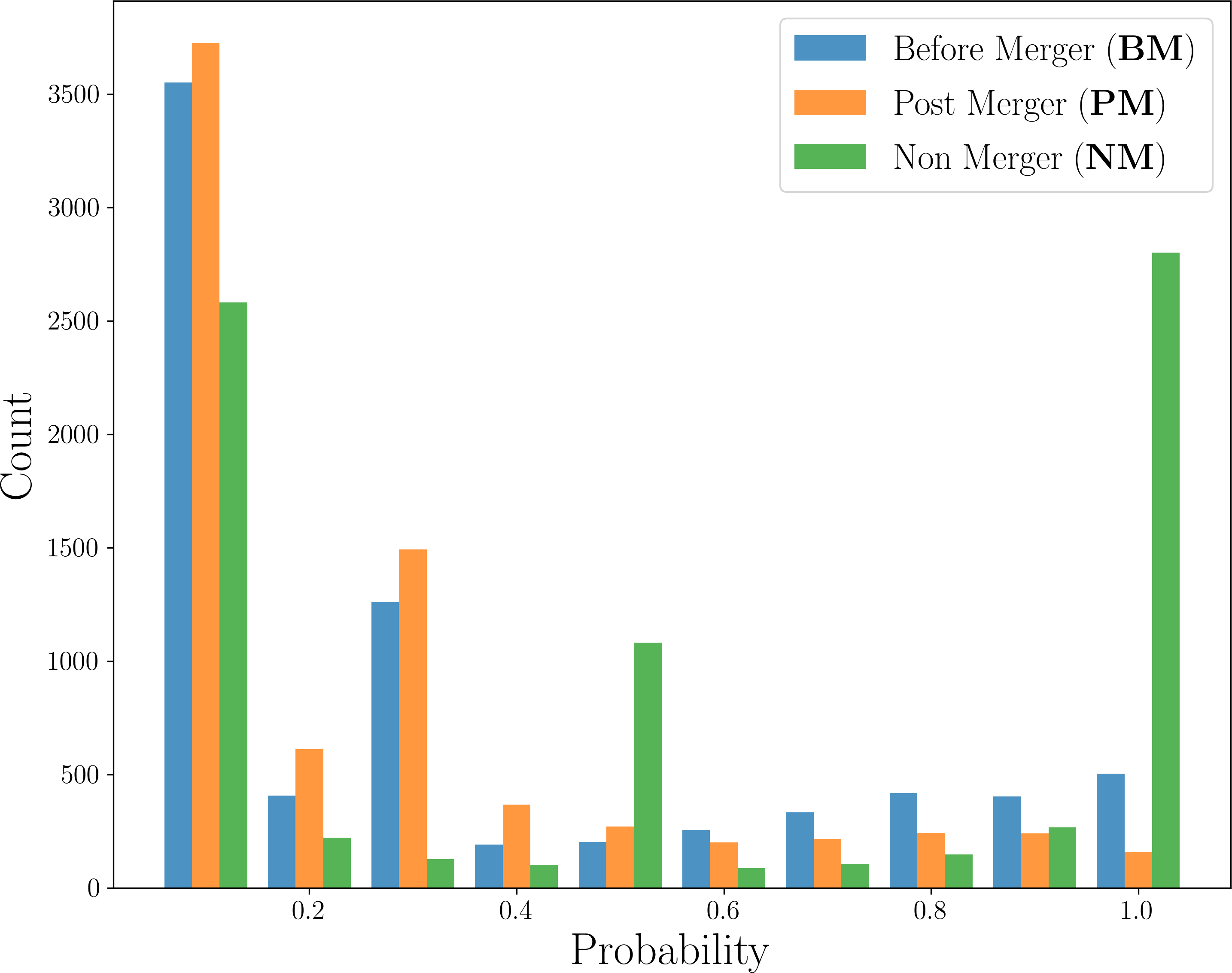}
    \caption{Probability distribution for the three classes that are classified by our hierarchy of models in the CANDELS selected sample. Overall these distributions are very distinct from the validation data. Here they are more irregular, especially those with intermediate confidence probabilities. This shows signs that the network is less certain about the classes in general than with was in the validation sample. This is expected since the validation sample is prepared to look very similar to but it is not equal to the CANDELS data.}
    \label{fig:prob_dist_candels}
\end{figure}

We test our methodology on CANDELS imaging data described in \S \ref{sec:CANDELSFIELDS}. For predicting classes on real data, we use an independent indicator to check if the observed galaxies are mergers or not. We rely on the visual classification of the CANDELS fields conducted in \cite{Kartaltepe2015}, where detailed information about the morphology is available. Using this, we have a set of indicators that can help us decide if the galaxy looks like a merger or not.  With this subsample of CANDELS galaxies that have similar properties to our simulation galaxies, we carry out predictions in the same way as we do for the validation data, as shown in Fig. (\ref{fig:prob_dist_candels}). {\color{black}{However, it is important to keep in mind that these visual indicators are not ground truths and are prone to the subjectivity of the classifiers. The apparent morphology of a galaxy merger can be produced by other physical processes.}}

%Our selection steps with CANDELS galaxies is outlined in Section (\ref{sec:CANDELSFIELDS}).
%We limit the CANDELS sample to the same stellar mass and redshift cuts used to do our selections in the simulation. We also filter the CANDELS catalogues to avoid images of stars, edges and other artifacts plus a magnitude cut of $H < 24.5 \ mag$ to ensure all images have reasonable signal-to-noise ratios.

\subsubsection{Visual Classification}

\begin{figure*}
    \centering
    \includegraphics[width=\textwidth]{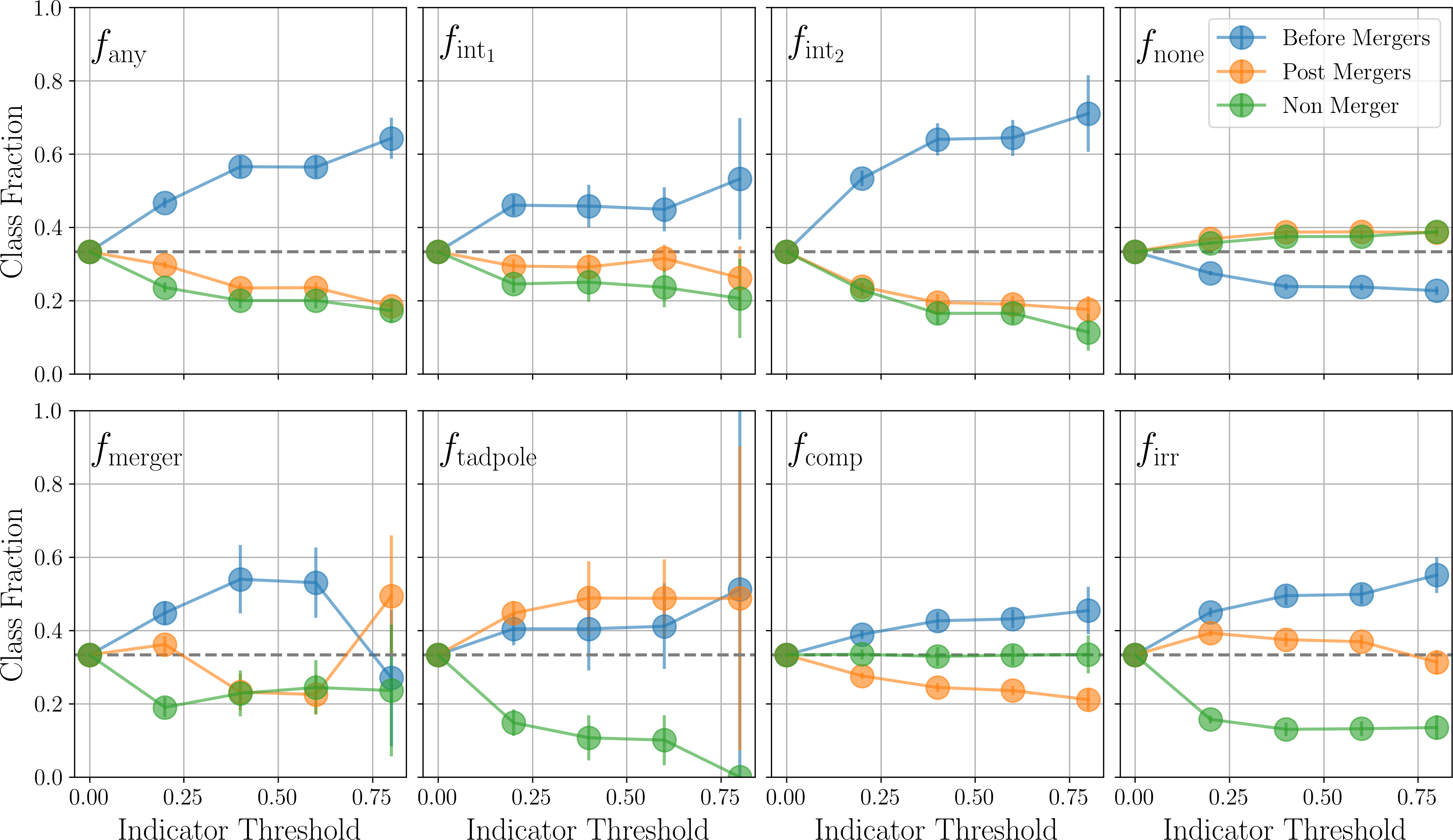}
    \caption{Mean class fractions from 100 samplings of a class balanced sub-sample (700 galaxies of each class) of CANDELS galaxies with the given indicator from visual classifications above the shown threshold. The first point represent the mean of the complete sub-sample of evenly distributed classes, while following points show only the fraction of those galaxies above the threshold. Error bars show $1 \pm \sigma$ for class fractions among all samples. \textbf{BM}, \textbf{PM} and \textbf{NM} are displayed in blue, orange and green, respectively.}
    \label{fig:visual_x_model}
\end{figure*}{}

The \cite{Kartaltepe2015} classification effort on CANDELS galaxies includes a set of indicators dedicated to describe galaxy mergers, with the goal to develop a group of characteristics only related to merging aspects of the morphology of the galaxy. Here, in order to assess how our model performs using real CANDELS galaxies, we compare how its classification relates to these indicators. 

Namely, we use the classification fractions \texttt{f\_any}, \texttt{f\_int1}, \texttt{f\_int2}, \texttt{f\_none}, \texttt{f\_merger}, \texttt{f\_comp}, plus two indicators that are not in the set of merger indicators but might relate to mergers, \texttt{f\_tadpole} and \texttt{f\_irr}. These fractions represent the overall fraction of total classifiers that marked the galaxy with given property. We briefly discuss each of these indicators here, for a full discussion please refer to \cite{Kartaltepe2015}.  

\texttt{f\_any} is used when the galaxy has any type of interaction. Usually, if a classifier marked a galaxy in any of the others indicators, it will also be marked with \texttt{f\_any}; \texttt{f\_int1} represent galaxies with interactions within their segmap, while \texttt{f\_int2} is for galaxies with interactions beyond their segmap; \texttt{f\_none} is used when the galaxy has no signs of interaction and \texttt{f\_merger} when the galaxy look like it underwent a recent merger event; \texttt{f\_comp} indicates if the galaxy has a non-interacting companion, with no signs of interaction and tidal features; The other two non-merger indicators, \texttt{f\_tadpole} and \texttt{f\_irr}, represents whether the galaxy look like a tadpole galaxy with strong tidal features, or if the galaxy has an irregular morphology, which in general might be a sign of merging, but not uniquely. So each indicator represents the fraction of classifiers that mark the galaxy as having the assigned characteristics. Thus, this fraction is related to how obvious and how unified the classification was among all expert classifiers. A fraction of 0 represents a galaxy that no classifier marked as having those characteristics, while a fraction of 1 represents the cases where all classifiers marked the galaxy with the given indicator. Intermediate fractions might result from morphologies that are ambiguous, thus objects with higher fractions represent less ambiguous morphologies. However, it is important to none that for some indicators very few objects were unanimously classified. Thus these indicators are subject to the subjectivity of the classifiers, while a higher fraction means that the classification is less prone to biases.

To explore how our model's classification of CANDELS galaxies correlates with the visual classification available from \cite{Kartaltepe2015}, we randomly generate 100 balanced sub-samples based on the model classification with 700 galaxies in each class. We do this as our resulting sample of CANDELS classified galaxies is very imbalanced towards non-mergers as shown in Fig. (\ref{fig:prob_dist_candels}). If we use the entire sample, trends in our class fraction would be more difficult to visualize, especially for the case of \textbf{PMs}, which consists of the class with the fewer number of classified objects. We then compare each sub-sample against increasing thresholds within  the given indicator. Fig. (\ref{fig:visual_x_model}) show the class fraction mean $\pm 1 \ \sigma$ for each class among all sub-sample for an increasing threshold. The \textbf{BMs} are shown in blue, \textbf{PMs} in orange and \textbf{NMs} in green. 

The overall trend with all merger indicators (\texttt{f\_any}, \texttt{f\_int1}, \texttt{f\_int2}, \texttt{f\_merger}) is dominated by an increase in the fraction of \textbf{BM} classifications, as one would expect. Plus, the fraction of \textbf{PMs} do not follow this trend with \textbf{BMs}, a sign that both classes represent different objects. Indeed, by solely following these merger indicators, one might assume that \textbf{PM} and \textbf{NM} represent the same type of objects since \texttt{f\_none} shows the fraction of \textbf{NM} and \textbf{PM} to be similar. However, \texttt{f\_tadpole} and \texttt{f\_irr} show similar trends for \textbf{BM} and \textbf{PM}. In this case, \textbf{PMs} classified by our model might represent galaxies without companions and clear signs of recent merger interactions by disturbed morphologies. Meanwhile, \texttt{f\_comp} show different behaviors for each class with a very small scatter, which suggest that \textbf{PM}s as classified by our network are isolated galaxies, with no clear signs of companions, while \textbf{NM} can have companions but no signs of interactions. This might represent a bias from the network towards objects without any companion in the field, which indicates that \textbf{BM} might have a significant impact from sky projections. On the other hand, this is expected since we do not factor in any redshift information in the central and neighbor galaxies in our classification method. The introduction of this information in the classification pipeline might further improve the quality of the model, but this is left for a future work.

\begin{figure*}
    \centering
    \includegraphics[width=\textwidth]{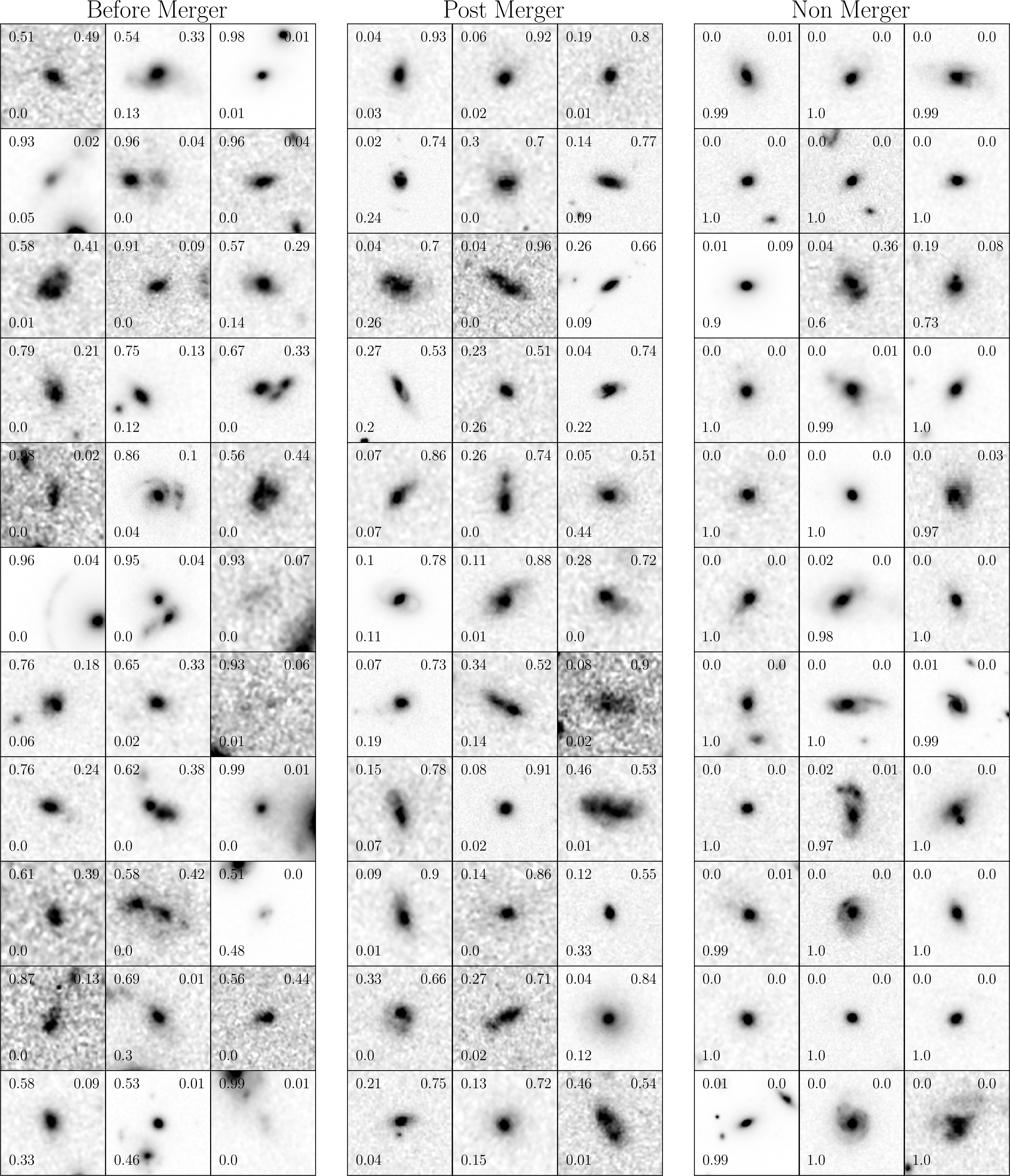}
    \caption{Mosaic with classifications done on CANDELS data for each class, \textbf{BM}, \textbf{PM} and \textbf{NM}, respectively. Mean probabilities for each class are shown in each image, top values represent merger classes (\textbf{BM} and \textbf{PM}) while bottom value represents the \textbf{NM} probability. The low probabilities represent cases where the network is more unsure and appears ambiguous. Increasing the probability threshold would produce more precise classifications with more clearly distinct morphologies, but we display here classifications above 50\% probability as this represents the peak completeness of our classifications and the threshold used throughout this paper. }
    \label{fig:CANDELS_MOSAIC}
\end{figure*}

In Fig. (\ref{fig:CANDELS_MOSAIC}) we show CANDELS galaxies as classified by our method with corresponding probabilities for each class, similarly to Fig. (\ref{fig:illustris_mosaic}). 

\subsubsection{Merger Fractions and Merger Rates}

One of our main goals in this paper is to estimate galaxy merger fractions, $f_{\rm m}$ and galaxy merger rates, $\mathcal{R}$, with our CNN method. We proceed to estimate $f_{\rm m}$ by counting merger classifications with probabilities $P(\rm class) > 0.5$ in $\Delta z = 0.5$ bins of redshift in the range $0.5 < z < 3$. We do this for both merger sub-classes, \textbf{BM}, \textbf{PM} and also for \textbf{MM}. Even though we train our model with low redshift galaxies, our CANDELS samples have only a few galaxies with redshifts $z < 0.5$, which results in poor statistics for merger fractions in that regime. The measured merger fractions we derive are shown in Table (\ref{tab:merger_fractions}).

We estimate galaxy merger rates by using merger fractions and appropriate timescales for each class, with $\tau_{\rm obs} = 0.3 \ \rm Gyr$ for \textbf{BM} and \textbf{PM}, and $\tau_{\rm obs} = 0.6 \ \rm Gyr$ for \textbf{MM}. Our timescales are defined by our sample selection steps, as described in \S \ref{subsec:major_mergers}. Although a consistent merger rate measurement does not validate individual classifications, it would represent that the overall statistics of the sample of classifications would follow one expected from other classification methods. By comparing merger rates estimated by our method with previous results we demonstrate a real application of our approach.

\begin{table}[h]
    \centering
    \begin{tabular}{cccc}
    \hline
    Redshift & BM & PM & MM \\
    \hline
    $0.5 \le z < 1.0$ & $0.041 \pm 0.008$ & $0.014 \pm 0.004$ & $0.055 \pm 0.009$  \\
    $1.0 \le z < 1.5$ & $0.048 \pm 0.009$ & $0.059 \pm 0.010$ & $0.107\pm 0.013$  \\
    $1.5 \le z < 2.0$ & $0.110 \pm 0.016$ & $0.084 \pm 0.014$ & $0.196\pm 0.021$  \\
    $2.0 \le z < 2.5$ & $0.180 \pm 0.032$ & $0.112 \pm 0.026$ &$ 0.292\pm 0.037$  \\
    $2.5 \le z < 3.0$ & $0.181 \pm 0.043$ & $0.206 \pm 0.044$& $0.383\pm 0.052$  \\
    \end{tabular}
    \caption{\textbf{BM}, \textbf{PM} and \textbf{MM} fractions in bins of redshift based on the classification from our models.}
    \label{tab:merger_fractions}
\end{table}

We estimate merger rates using our model by simply taking our merger fractions averaged over our timescale, that is

\begin{equation}
    \mathcal{R} = \frac{f_{m}}{\tau_{\rm obs}}.
\end{equation}{}

\begin{figure*}
    \centering
    \includegraphics[width=\textwidth]{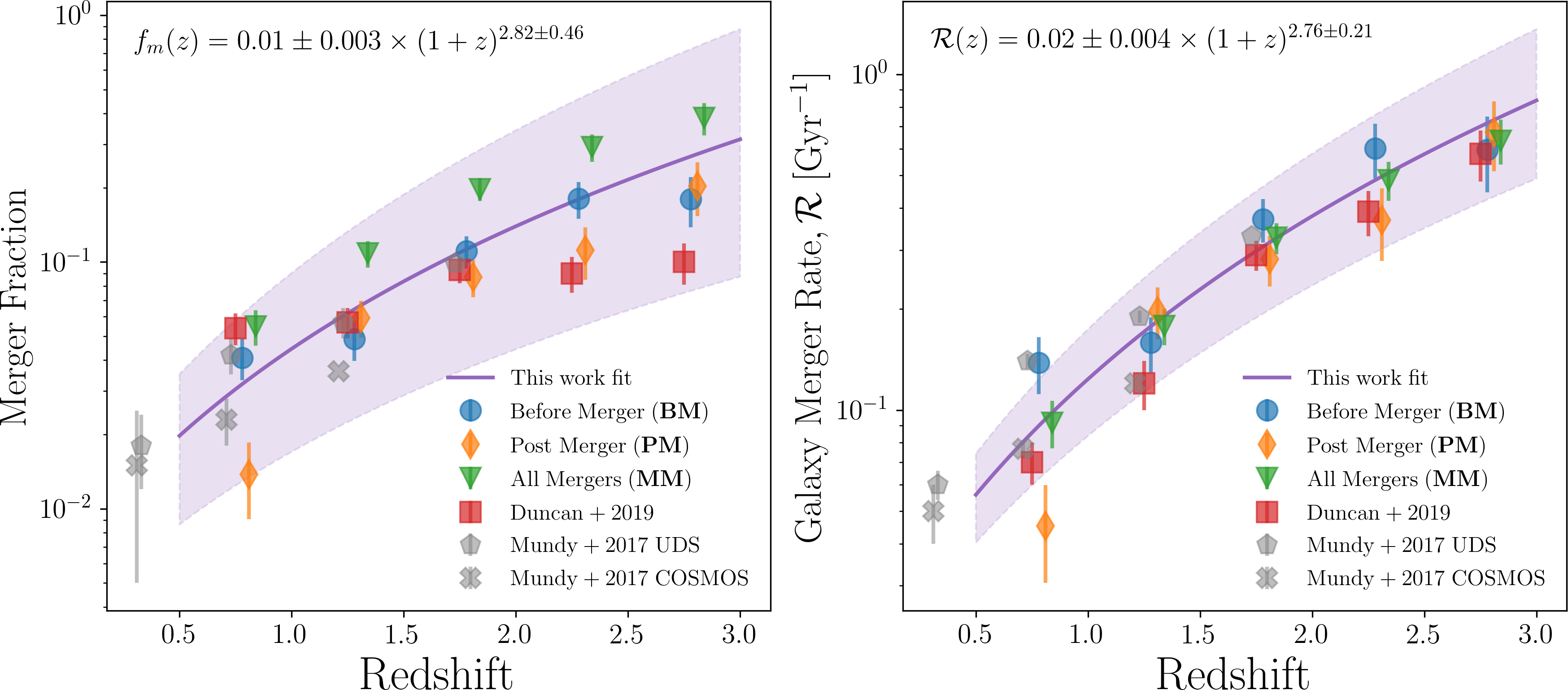}
    \caption{Merger fractions $f_m$ (left) and galaxy merger rates $\mathcal{R}$ (right) in bins of redshift for our \textbf{BM} (blue circles), \textbf{PM} (orange diamonds) and \textbf{MM} (green triangles) classifications. Error bars represent $\pm 1 \ \sigma$ uncertainties and account for the accuracies displayed in the confusion matrix in Fig. (\ref{fig:cm_illustris}). We fit a power law for fractions and rates and show the best fit in purple together with $\pm 1 \ \sigma$ uncertainties of the fit in the shaded area. We show results from \cite{Duncan2019} (red squares) and from \cite{Mundy2017} (gray X's and hexagons) for comparison. Overall, the trend estimated by our model agrees very well with previous results. Best fitting parameters and uncertainties are shown in the upper left corner of both plots. }
    \label{fig:merger_frac_rates}
\end{figure*}

We plot our estimated merger fractions and rates in Fig. (\ref{fig:merger_frac_rates}), in the left panel and right panel respectively, comparing with the results of merger fractions and rates as estimated with CANDELS galaxies from \cite{Mundy2017} and \cite{Duncan2019}. 

One important point is that our model was not prepared to measure merger fractions by construction, as it was trained with a balanced sample of mergers and non-mergers. Additionally, no redshift bias for mergers was used. In fact, the redshift distribution of our training sample is also balanced between mergers and non-mergers (Fig. \ref{fig:redshiftdist}).

It is possible to check in Fig. (\ref{fig:merger_frac_rates}) that our results are in general consistent with merger rates found by \cite{Mundy2017} and \cite{Duncan2019}. Here, even though we are making comparisons to close pairs statistics results, we do not make any assumptions on the fraction of pairs that will actually merge, $C_{pair}$, in $\mathcal{R}$ as all galaxies considered as mergers in our training sample are actually mergers, as we use information from IllustrisTNG's merger trees. Moreover, based on our selection approach, we are also not introducing information about the simulation's intrinsic merger rates into our model. 

We fit power laws to our merger fractions and rates of the form
\begin{equation}
    f_m(z) = f_0 \times (1+z)^m
\end{equation}{}
\begin{equation}
    \mathcal{R}(z) = \mathcal{R}_0 \times (1+z)^m,
\end{equation}{}
\noindent to our merger fractions and rates respectively. We do this fit by a simple least squares fit to all our data points, including \textbf{BM}, \textbf{PM} and \textbf{MM}, and show the uncertainty based on $\pm 1 \ \sigma$ (shaded region in Fig. \ref{fig:merger_frac_rates}). We find

\begin{equation}
    f_m(z) = 0.01 \pm 0.003 \times (1+z)^{2.82 \pm 0.46},
\end{equation}{}
and
\begin{equation}
    \mathcal{R}(z) = 0.02 \pm 0.004 \times (1+z)^{2.76 \pm 0.21},
\end{equation}{}

\noindent which is expected since our observing timescale, $\tau_{\rm obs}$, is flat and defined by our selection (\S \ref{subsec:major_mergers}). Overall this shows that the trend represented by our findings using major merger classifications by a deep learning model agrees with the trend found by \cite{Duncan2019} using close pair statistics for all the CANDELS fields, where within the redshift range probed here $0.5 < z < 3$, the highest merger rates, $\mathcal{R}$, are found in the highest redshift probed. Different assumptions regarding timescales and a different method of identifying mergers yield similar results, and even though our uncertainty is larger at all redshifts, the mean of our classifications match pairs well.

We cannot probe higher redshifts with our current model as it is limited by our training data, which was prepared to probe redshifts up to $z = 3$ with observed near-infrared data. One could expand the model to probe higher redshifts by training it with rest-frame UV data, but in this case the effects of dust and the lack of a radiative transfer treatment would become more important and the training sample should be prepared in a different manner, however this will be examined in a future study.

\section{Summary} \label{sec:summary}

In this work we show that it is possible to train deep learning models to find galaxy mergers using only simulated galaxies and then to carry out predictions on real data by training a deep learning Convolutional Neural Network (CNN) model. We do this by classifying galaxy mergers with IllustrisTNG data and then carrying out predictions on real CANDELS galaxies. We show that

\begin{itemize}

    \item Using automated methods for optimizing deep learning hyperparameters is a good way of achieving high performance architectures for solving astronomy classification tasks. This not only speeds up the training step of working with deep learning networks, but removes some of the subjectivity present when fine tuning such hyperparameters by hand.

    \item It is possible to train a model capable of achieving $\sim$90\% accuracy in classifying galaxy mergers within the simulated balanced validation sample. Not only that, but our model can classify mergers in two stages: mergers before the merger event (\textbf{BM}) and post mergers \textbf{PM}, with 87\% and 78\% accuracy, respectively. The performance of the model using simulated galaxies from IllustrisTNG does not directly translate to the same performance that would be achieved using real galaxies, as the validation sample is balanced in the simulation, which is not true in our CANDELS sample. The quality of the model with real galaxies must be assessed by the visual classification comparison and the estimated galaxy merger rates.
    
    \item We show that predictions using real galaxy images are possible, and galaxies classified in the validation and CANDELS samples share similarities. We show that our model classifications follows visual classification indicators for mergers from \cite{Kartaltepe2015}. Even though merger classifications can be ambiguous between visual classifiers, our blind classifications based on the information from mergers trees from the IllustrisTNG show that galaxy mergers classified by our network have similar visual cues to those classified by visual experts. This is shown by the different trends for mergers before the merger event, post mergers and non-mergers when compared to merger indicators from visual classifications. Galaxies before the merger event (\textbf{BM}) dominate samples selected with higher thresholds of the merger indicators from the visual classification. 
    
    \item By using our model to classify CANDELS galaxies we measure galaxy merger fractions and rates between $0.5 \le z \le 3$ that are consistent with previous results for CANDELS galaxies estimated with close pair statistics from \cite{Duncan2019}. This was done without any prior merger fraction or rate information embedded in our training step. Our model, by construction, was not prepared to do such measurements and this is an independent method of estimating merger fractions and rates, even though the uncertainties are higher than when using other methods. 
    
\end{itemize}{}

Our results are based on a sample of simulated galaxies with several constraints: our mocks do not account for the effects of dust, we do not explore arbitrary orientations besides face-on and edge-on orientations, and our results are only limited to massive galaxies with $M_* > 10^{10} M_\odot$. Addressing these points will further improve results when carrying out predictions on real galaxies, as it would serve to lessen the gap between simulated and real galaxies. This approach is limited by the quality of the training data, and improvements in the post-processing of the simulation data should further improve the results displayed here. It is of utmost importance to always use large training samples, as the parameter space in the training step is crucial for the learning of the model.

This work shows the potential of using a combination of galaxy simulations and machine learning techniques as an avenue for solving problems where observables are impossible or expensive to estimate from real observations of galaxy mergers. Approaches like the one presented here will naturally improve alongside cosmological simulations.

\section{Acknowledgments}

The authors would like to thank the anonymous referee for their suggestions and comments that led to significant improvements on the paper and the Centre for Astronomy and Particle Theory of University of Nottingham for providing all computational infrastructure necessary to run the training steps to produce the model described here. This study was financed in part by the Coordena\c{c}\~{a}o de Aperfei\c{c}oamento de Pessoal de N\'{i}vel Superior - Brazil (CAPES). KJD acknowledges support from the ERC Advanced Investigator programme NewClusters 321271. TYC acknowledges the support of the Vice-Chancellor's Scholarship from the University of Nottingham. AG and AW acknowledges funding from the Science and Technology Facilities Council (STFC).

\software{Astropy \citep{2018AJ....156..123A},
Matplotlib \citep{2007CSE.....9...90H}, Morfometryka \citep{Ferrari2015}, Scikit-Learn \citep{scikit-learn}}




\bibliography{referencesmend}



\end{document}